\documentclass{aa}  

\usepackage{graphicx}
\usepackage{txfonts}
\usepackage{lipsum}
\usepackage{subcaption}         
\usepackage{lscape}             
\usepackage{placeins}           
\usepackage{csquotes}
\usepackage{hyperref}
\usepackage{tabularx}
                                
\begin{document}

   \title{Dynamical mirages: how bar-induced resonant trapping can mimic substructure clustering in dynamical parameter spaces}


   \author{M. De Leo\inst{1,2}\thanks{micheledl89@gmail.com}
        \and D. Massari\inst{2}
        \and M. Bellazzini\inst{2}
        \and A. Mucciarelli\inst{1,2}
        \and B. Acosta-Tripailao\inst{3,4}
        \and C. Nipoti\inst{1}
        }

   \institute{Dipartimento di Fisica e Astronomia, Universit\`{a} degli Studi di Bologna, Via Piero Gobetti 93/2, Bologna, 40129, Italy
            \and Osservatorio di Astrofisica e Scienza dello Spazio di Bologna, INAF, Via Piero Gobetti 93/3, Bologna, 40129, Italy
            \and Instituto de Astrof\'{i}sica, Pontificia Universidad Cat\'{o}lica de Chile, Av. Vicu\~{n}a Mackenna 4860, 782-0436, Macul, Santiago, Chile           
            \and Instituto Milenio de Astrof\'{i}sica MAS, Av. Vicu\~{n}a Mackenna 4860, 782-0436, Macul, Santiago, Chile}

   \date{Received xxx}

 
  \abstract
   {The complex task of unraveling the assembly history of the Milky Way is in constant evolution with new substructures identified continuously. To properly validate and characterise the family of galactic progenitors, it is important to take into account all the effects that can shape the distribution of tracers in the Galaxy. First among the often overlooked actors of galactic dynamics is the rotating bar of the Milky Way that can affect orbital tracers in multiple ways.}
   {We want to fully characterise the effect of the rotating bar of the Milky Way on the distribution of galactic tracers, provide diagnostics helpful in identifying its effect and explore the implications for the search and identification of substructures.} 
   {We use the in-house Orbital Integration Tool (\textsc{OrbIT}), built to include the full effect of the bar and exploit its multidimensional output to perform a complete dynamical characterisation of a large sample of carefully selected Milky Way stars with very precise astrometry.}
   {We identify conspicuous overdensities in several orbital parameter spaces and verify that they are caused by the bar-induced resonances. We also show how contamination by trapped tracers provides local density enhancements that mimic the clumping usually attributed to genuine substructures.}
   {We provide a new and expedite way of identifying resonant loci and, consequently, to estimate the contribution of stars trapped into orbital resonances to phase-space overdensities previously identified as candidate relics of past merging events. Among those analysed here, we found that the detections of Cluster 3 and Shakti seem to have gained a non-negligible boost from resonance-trapped stars. Nyx is the most extreme case, with $\sim70\%$ of assigned member stars lying on resonant orbit, strongly suggesting that, in fact, it is not the genuine relic of a merger event but, instead, an overdensity caused by bar-induced resonances.
   }


   \titlerunning{Dynamical mirages: bar-induced resonant trapping mimicking substructures}
   \maketitle

\section{Introduction}\label{intro}
Study of the evolution and structure of the Milky Way (MW) is continuously ongoing. The premier technique used to identify structures, moving groups and candidate relics of ancient merger events is to look for overdensities and clumps in Integrals of Motion (IoMs) spaces \citep[chiefly the angular momentum vs energy space, $L_z-E_{tot}$][]{2000MNRAS.319..657H} or in the action spaces \citep[i.e.][]{2018ApJ...856L..26M}. Recovery of the mentioned orbital parameters is model dependent, as it is based on the underlying potential used to reconstruct the orbital history of the tracers. The technique of \textquote{orbit integration} sits at the crossroad of observations and theory as it uses observed starting positions and velocities of tracers and the theoretical model of a density distribution to compute the forces dictating the motions of the tracers and reconstruct their orbits. Modeling of structure in simulations is following closely observational evidence, with complex multicomponent potential distributions aiming to reproduce the density of both the observed luminous matter and the inferred dark matter (DM). The same cannot be said for the models used for orbital integration, with most studies still conducted using underlying static, axisymmetric potential models. These remove the ability to identify the effects deriving from a time-varying potential, such as the secular evolution of the MW and the effect of its rotating bar. The latter in particular has a multifaceted influence on the dynamics of galactic tracers, from the non-conservation of some IoMs to the shepherding and trapping of tracers in resonances. Theoretical predictions of the action of the bar are plentiful in the literature, starting from the effect it could have on the disc \citep{1991dodg.conf..323K, 2013MNRAS.430.3276M, 2021MNRAS.500.4710C, 2021MNRAS.505.2412C}, on the stellar halo \citep{2015MNRAS.451..705M, 2021MNRAS.506.4687M} and on the DM halo \citep{1998ApJ...493L...5D, 2000ApJ...543..704D}. One of the first examples of a peculiar moving group linked to the effect of the bar was the Hercules group \citep{1958MNRAS.118..154E, 2000AJ....119..800D, 2017ApJ...840L...2P, 2020ApJ...890..117D}. Initially associated with the Outer Lindblad Resonance under models of a fast rotating, small bar \citep[i.e.][with $\Omega_p = 55 \ {\rm km \ s^{-1} \ kpc^{-1}}, a_{bar} = 3 \ {\rm kpc}$]{2000AJ....119..800D}, later studies have shown it is trapped around the $4^{th}$ Lagrangian point, a stable region along the bar minor axis, around the corotation resonance \citep[i.e.][with $\Omega_p = 40 \ {\rm km \ s^{-1} \ kpc^{-1}}, a_{bar} = 4.5 \ {\rm kpc}$]{2020ApJ...890..117D}. The advent of the \textit{Gaia} catalogues \citep{2016A&A...595A...2G}, with their unprecedented volume of data, opened up the possibility of detailed and statistically meaningful analysis of stellar kinematics. This led to a more systematic study of the effect of the bar-induced resonances on tracer populations \citep{2020A&A...634L...8K, 2022A&A...663A..38K, 2023MNRAS.524.3596D, 2024ApJ...971L...4D, 2024MNRAS.532.4389D}. At the same time, more sophisticated clustering techniques and algorithms led to the identification of new substructures \citep[i.e. for some of the latest developments][]{2022A&A...665A..57L, 2022A&A...665A..58R, 2023A&A...670L...2D, 2024ApJ...976..161L}. The debate on how to validate substructures (both new and previously identified) with dynamical information is in full swing, with very recent works \citep{2025A&A...700A.240W, 2025MNRAS.542.1331D} warning of the dangers of employing clustering algorithms on the $L_z-E_{tot}$ plane without accounting fully for the bar. One key question that remains unanswered is if the bar-induced resonances can affect the distribution of tracers in dynamical phase spaces profoundly enough to contaminate or even mimic the presence of a real substructure.

With the aim of studying the full effect of the rotating bar on the dynamics of MW tracers, we explore the dynamical parameter space of a large sample of MW stars with exquisite astrometry \citep[][fully presented in Sec.~\ref{data}]{2023A&A...674A.194B}. We conduct our study with the \textsc{OrbIT} code (De Leo et al. submitted), developed to fully account for the effect of the rotating bar and described, together with its underlying potential model, in Section~\ref{method}. Section ~\ref{results} contains our findings for the main sample while in Sec.~\ref{subs} we extend our study to a selection of known substructures of the MW. We discuss our results in Section ~\ref{discu} and summarise them in our conclusions (Sec.~\ref{conclusion}).

\section{Data}\label{data}
The main sample analysed in this work is presented in \citet{2023A&A...674A.194B} and available as a public catalogue \citep{2023yCat..36740194B}\footnote{\href{https://cdsarc.cds.unistra.fr/viz-bin/cat/J/A+A/674/A194}{$https://cdsarc.cds.unistra.fr/viz-bin/cat/J/A+A/674/A194$}}. Briefly, the sample is composed of 694233 giants from the \textit{Gaia} Synthetic Photometry Catalogue \citep[GSPC, ][]{2023A&A...674A..33G} selected excluding probable binary, variable and heavily extincted sources and non-genuine RGB stars. For the purposes of our work, the main characteristic of the catalogue is the exquisite astrometry deriving from the quality cuts on \texttt{parallax\_over\_error} $> 10$, \texttt{RUWE} $< 1.3$ \citep{2021A&A...649A...2L} and $\simeq 90\%$ of the stars with radial velocity measurements with a precision better than $3.0 \ {\rm km/s}$. These selections ensure a high standard of precision in all phase space coordinates ($\alpha, \ \delta, \ {\rm D}, \ \mu_{RA}, \ \mu_{Dec}, \ {\rm RV}$), with the quality of the parallaxes ensuring precise recovery of both distance and proper motion measurements.

\section{Method}\label{method}

This work is conducted mainly using the custom-made orbit integration code \textsc{OrbIT}, presented and detailed in De Leo et al. (submitted). Briefly, \textsc{OrbIT} is a code for orbit integrations written in the C computing language and using a \textquote{kick-drift-kick} integration scheme \citep[a modified type of Verlet integration scheme, ][]{1967PhRv..159...98V,2008gady.book.....B} that is time-reversible and suppresses numerical errors \citep{1992PhR...216...63R,2003AcNum..12..399H}. For our purposes, the main and most important difference between \textsc{OrbIT} and other publicly available codes like \textsc{galpy} \citep{2015ApJS..216...29B} or \textsc{AGAMA} \citep{2019MNRAS.482.1525V} lies in the way it accounts for the time variability of the potential underlying the integration. While the codes mentioned above offer the option of including the rotating bar of the MW in the potential, they still work within a spherically symmetric theoretical framework. This means that they do not account for the fact that the time variability of the barred potential implies that each individual radial oscillation (i.e. a single passage from pericentre to apocentre) of each tracer is slightly different from the previous ones. This makes it difficult to gauge the precision of the orbital parameters given in output as a single value. For this reason, \textsc{OrbIT} computes one value of $R_{peri} \ {\rm and} \ R_{apo} \ {\rm (and \  therefore} \ ecc {\rm )}$ for each individual orbit for each tracer and the final output values are the mean of these distributions, with the standard deviations as associated errors. The quality of the selected sample and the statistical nature of our analysis ensure that the observational uncertainties do not have an impact on our results, nevertheless we include a test with maximal errors in Sec.~\ref{results}.

While theoretically there is no obstacle to computing the orbital parameters in this way with \textsc{galpy} and \textsc{AGAMA} (as shown in the example with \textsc{AGAMA} in Fig.~\ref{fig:agama} and discussed in Sec.~\ref{results}), it would require the users to step back and analyse on their own the \textquote{raw} orbits produced by the codes. Thus the strength of \textsc{OrbIT} with respect to the other available codes mostly lies in readily providing the outputs needed to conduct the present study.

To provide \textsc{OrbIT} with starting positions and velocities for the tracers, we transformed the observed positions, distances, proper motions and radial velocities to the Galactocentric reference frame. The transformation is done assuming the velocity vector of the Sun to be $(U_{\odot}\,,V_{\odot}\,,W_{\odot})=(11.1\,,12.24\,,7.25) \ {\rm km \ s^{-1}}$ \citep{2010MNRAS.403.1829S}, the velocity of the Local Standard of Rest $V_{LSR}=232.8 \ {\rm km \ s^{-1}}$ \citep{2017MNRAS.465...76M}, the distance of the Sun from the Galactic Center $R_{\odot}=8.20 \ {\rm kpc}$ \citep{2019A&A...625L..10G}, and the height of the Sun above the Galactic Plane $z_{\odot}=14 \ \rm{ pc}$ \citep{1997MNRAS.288..365B}.

Finally, the orbital history of the tracer stars examined in this work is computed with \textsc{OrbIT} backwards in time for a total of 5 Gyr with a timestep of $10^3 \ $ yrs.

\subsection{Galactic potential model}\label{MWpot}

The MW potential underlying the \textsc{OrbIT} integrations is a time-varying model including several components, for a detailed description see De Leo et al. submitted. Briefly, the MW is represented by a Navarro-Frenk-White \citep[NFW, ][]{1996ApJ...462..563N} Dark Matter (DM) halo, four Miyamoto-Nagai \citep[MN, ][]{1975PASJ...27..533M} discs (two stellar and two gaseous), and a bulge composed of a Long-Murali \citep[LM, ][]{1992ApJ...397...44L} rotating bar and a Plummer \citep{1911MNRAS..71..460P} spherical component. The parameters for the disc are fit to recover the respective local mass densities \citep[i.e. ][]{2015ApJ...814...13M, 2017MNRAS.465...76M, 2022MNRAS.513.4130L} and the bar is geometrically similar to the one from \citet{2015MNRAS.450.4050W}, inclined at $\alpha_{bar} = 30^{\circ}$ and rotating with a pattern speed $\Omega_p = 41.3$\,km\,s$^{-1}$\,kpc$^{-1}$ \citep{2019MNRAS.488.4552S}. The full list of parameters of the different components of the MW potential is reported in Table~\ref{tab_pot}.

\begin{table}
 \caption{Parameters of the components of the MW potential model used. Sources: 1 = \citet{2015ApJS..216...29B}, 2 = \citet{2017MNRAS.465...76M}, 3 = \citet{2022MNRAS.513.4130L}, 4 = \citet{2015ApJ...814...13M}, 5 =  \citet{2015MNRAS.450.4050W}, 6 = \citep{2016A&A...587L...6V}, 7 = \citep{2018A&A...618A.147Z}.}\label{tab_pot}
  \begin{tabular}{| c c c c c |}
 \hline
 DM halo & M ${\rm [M_{\odot}]}$ & ${\rm r_s}$ ${\rm [kpc]}$ & c & Sources \\
 \hline
   & $8 \times 10^{11}$ & 16.0 & 15.3 & 1 \\
 \hline
 Discs & M ${\rm [M_{\odot}]}$ & ${\rm a \ [kpc]}$ & ${\rm z \ [kpc]}$ & Sources \\
 \hline
 Thin & $3.65 \times 10^{10}$ & 3.5 & 0.3 & 2, 3 \\
 Thick & $1.55 \times 10^{10}$ & 2.0 & 0.9 & 2, 3 \\
 Gas I & $1.1 \times 10^{10}$ & 1.824 & 0.085 & 2, 4 \\
 Gas II & $1.2 \times 10^{9}$ & 5.895 & 0.045 & 2, 4 \\
 \hline
 Bulge & M ${\rm [M_{\odot}]}$ & ${\rm r/a \ [kpc]}$ & ${\rm b, \ c \ [kpc]}$ & Sources \\
 \hline
 Bar & $1.0 \times 10^{10}$ & 5.5 & 0.68, 0.09 & 5 \\
 Spheroidal & $1.0 \times 10^{10}$ & 0.3 & - & 6, 7 \\
 \hline
 \end{tabular}
\end{table}

\subsection{Outputs}\label{out}

\textsc{OrbIT} can be customised to have a wide variety of outputs. In its most common form it provides all the classical orbital parameters, IoMs, the actions and other adiabatic invariants, and the circularity of each tracer. It can also output a full orbital history recording the 6D spatial information of each tracer at every timestep. For the present work we are mainly interested in the classical orbital parameters $R_{apo}, \ R_{peri}, \ ecc$, the actions $J_{\phi} = L_z \ {\rm and} \ J_R$ and the \textquote{characteristic} IoMs of energy, $E_{cha}$, and angular momentum, $L_{z,cha}$, \citep{2015MNRAS.451..705M, 2021MNRAS.506.4687M}. The characteristic IoMs are defined as the mean between the maximum and minimum value of the respective IoM during the course of the entire integration while the action $J_R$ is computed as in \citet{2008gady.book.....B}:
\begin{equation}\label{Jr}
    J_R = \frac{2}{\pi}\int_{R_{peri}}^{R_{apo}} \sqrt{2E-2\widetilde{\Phi}(r)-\frac{L^2}{r^2}}\rm{d}r \ ,
\end{equation}
where $E$ and $L$ are respectively the initial energy and total angular momentum of the tracer, $\widetilde{\Phi}(r)$ is a numerical approximation of the potential (including all components) affecting the tracer at distance $r$ and the distances varies between the $R_{apo} \ {\rm and} \ R_{peri}$ found by the orbit integration. In $\widetilde{\Phi}(r)$, the non-spherical components of the potential (discs and bar) are approximated with sphericalised forms (i.e. depending only on $r$). The approximations minimise the differences with the corresponding analytical potential when taking the mean of all axial components. The integral is solved numerically using a composite Simpson's 1/3 rule \citep{1991numan.book.....A} and the results are comparable with those derived from the St\"{a}ckel-Fudge method (see App~\ref{Jr_comp}).

The orientation of the Galactocentric reference frame axes implies that prograde tracers have negative $L_z \ {\rm and} \ J_{\phi}$ while retrograde tracers have positive values of these parameters.

\begin{figure}
\centering
\includegraphics[width=\hsize]{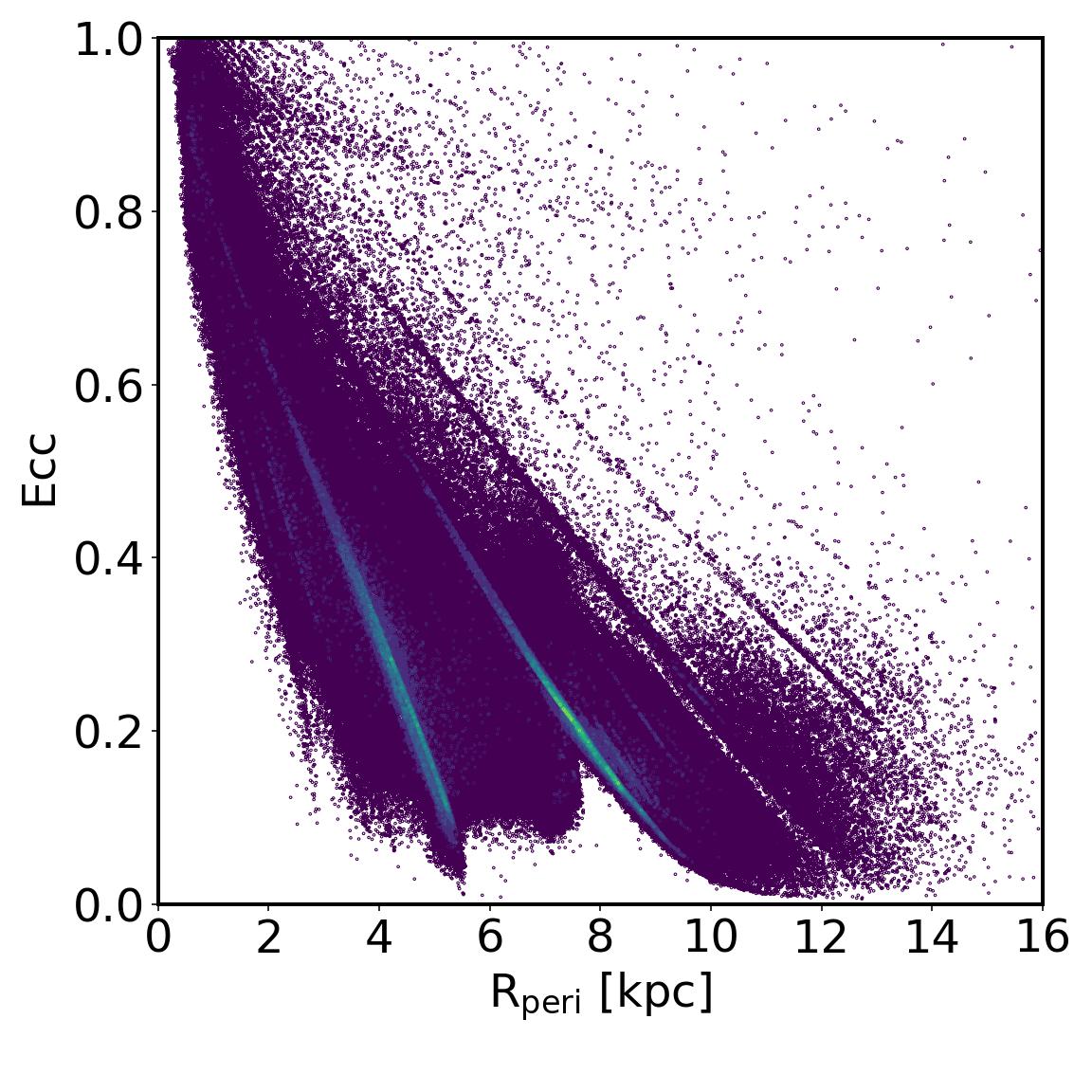}
\caption{Density distribution of the sample stars in $R_{peri}-ecc$ plane, higher density is indicated by a lighter colour.}
    \label{fig:periecc}
\end{figure}

\begin{figure}
\centering
\includegraphics[width=\hsize]{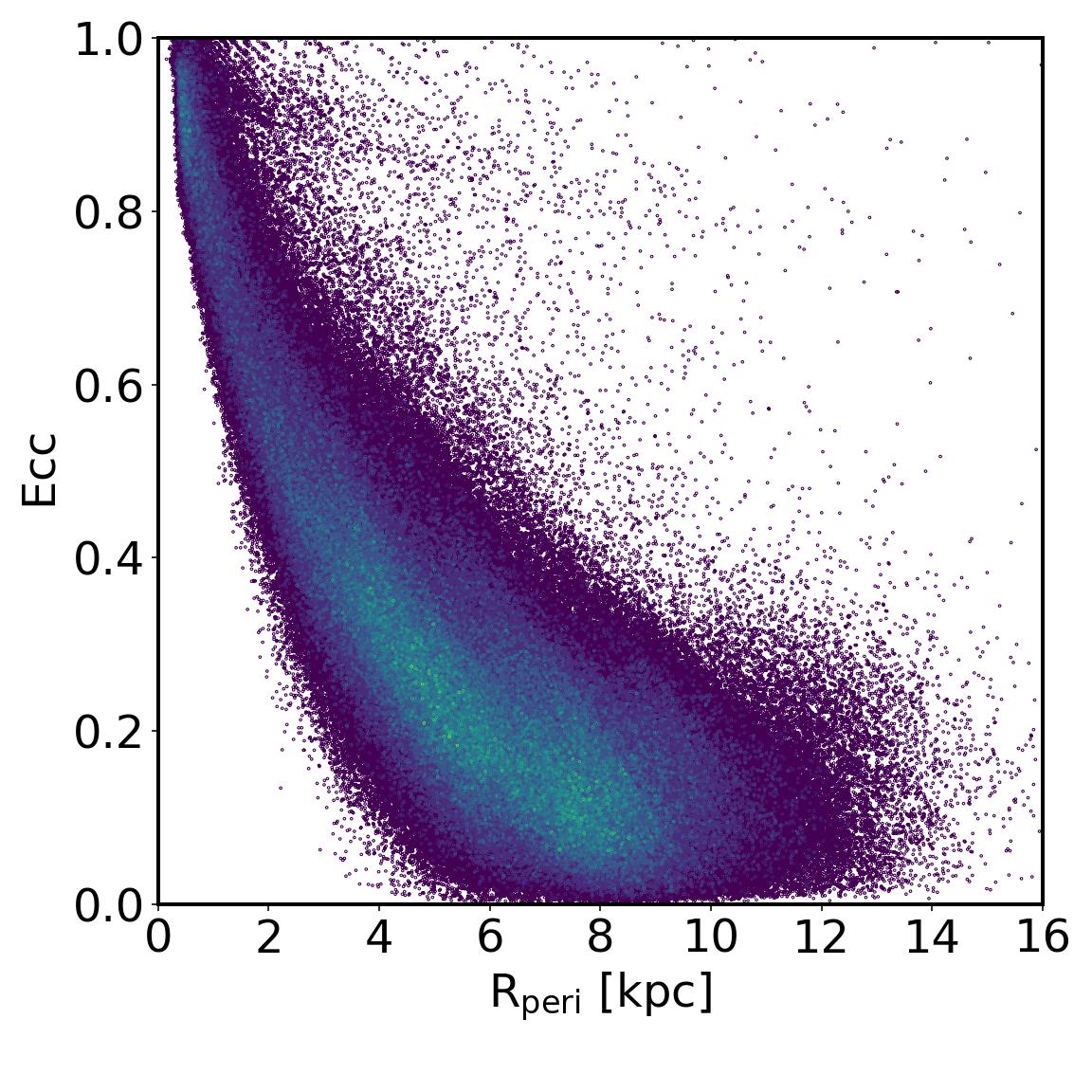}
\caption{Density distribution of the sample stars in $R_{peri}-ecc$ plane for an integration with a static axisymmetric potential (without the rotating bar). There is no trace of the overdensities seen in Fig.~\ref{fig:periecc}.}
    \label{fig:periecc_nobar}
\end{figure}

\section{Bar-induced overdensities in orbital phase spaces}\label{results}

The sample of \citet{2023A&A...674A.194B} allows to study in detail large regions of various dynamical parameter spaces. The most notable features emerging from our analysis of the \textsc{OrbIT} outputs are a series of overdensities clearly visible in the spaces of the orbital parameters, $R_{peri}, \ R_{apo} \ {\rm and} \ ecc$. Figure ~\ref{fig:periecc} shows the clearest case of these overdensities appearing as diagonal sequences in the $R_{peri}-ecc$ plane.

First, we checked the effect of including observational uncertainties. Due to the strict quality selection of our dataset and the statistical nature of the results (the presence of the overdensities), we didn't expect that including the small observational errors would make a difference. We performed a test by changing the input values by a symmetric $\pm1\sigma$, and verified that the overdensities endured and that we recovered similar amounts of tracers within them: $30.36\%$ in the nominal case, $29.74\% {\rm \ and } \ 30.87\%$ for the two extreme cases with the errors. Thus the observational errors, while they might change the result for the individual tracer, have a negligible impact on the statistical properties of the population.

We then performed an extensive range of tests to ensure that these features were not spurious effects due to errors in the orbital integration code, or in the (rather straightforward) analysis. Among other things, we reran the integration under various conditions, changing: the components of the potential model, the integration time step, the total integration time, the bar pattern speed, and the \textquote{direction} of the integration (going forward in time). Most notably among these tests we ran one with a static and axisymmetric version of the potential without the bar (but conserving the total bulge mass) and another version with a non-rotating bar. These tests confirmed that the overdensities were present only in the tests including a rotating bar and disappeared if the bar was absent or non-rotating (see Fig.~\ref{fig:periecc_nobar} for the case with the static and axisymmetric potential without the rotating bar). The array of tests highlighted that the pattern speed is the single parameter with the greatest impact on the prominence, shape and location of the overdensities. For pattern speeds between $35 \ {\rm and} \ 45 \ {\rm km \ s^{-1} \ kpc^{-1}}$ the overdensities are evident while for values outside this range the overdensities get progressively less populated (starting from the most eccentric tracers). On the low end of pattern speed values, we found some overdensities still recognisable at $20 \ {\rm km \ s^{-1} \ kpc^{-1}}$ while they were erased at $10 \ {\rm km \ s^{-1} \ kpc^{-1}}$. On the high end, it is possible to identify the overdensities up to $80 \ {\rm km \ s^{-1} \ kpc^{-1}}$ but mostly for tracers with $ecc \leq 0.5$. In Appendix~\ref{ps35} we show how our results are robust to changes in the pattern speed by carrying out the entire analysis with $\Omega_p = 35 \ {\rm and} \ 45 \ {\rm km \ s^{-1} \ kpc^{-1}}$.

\begin{figure*}
     \centering
     \begin{subfigure}[b]{0.48\textwidth}
         \centering
         \includegraphics[width=\textwidth]{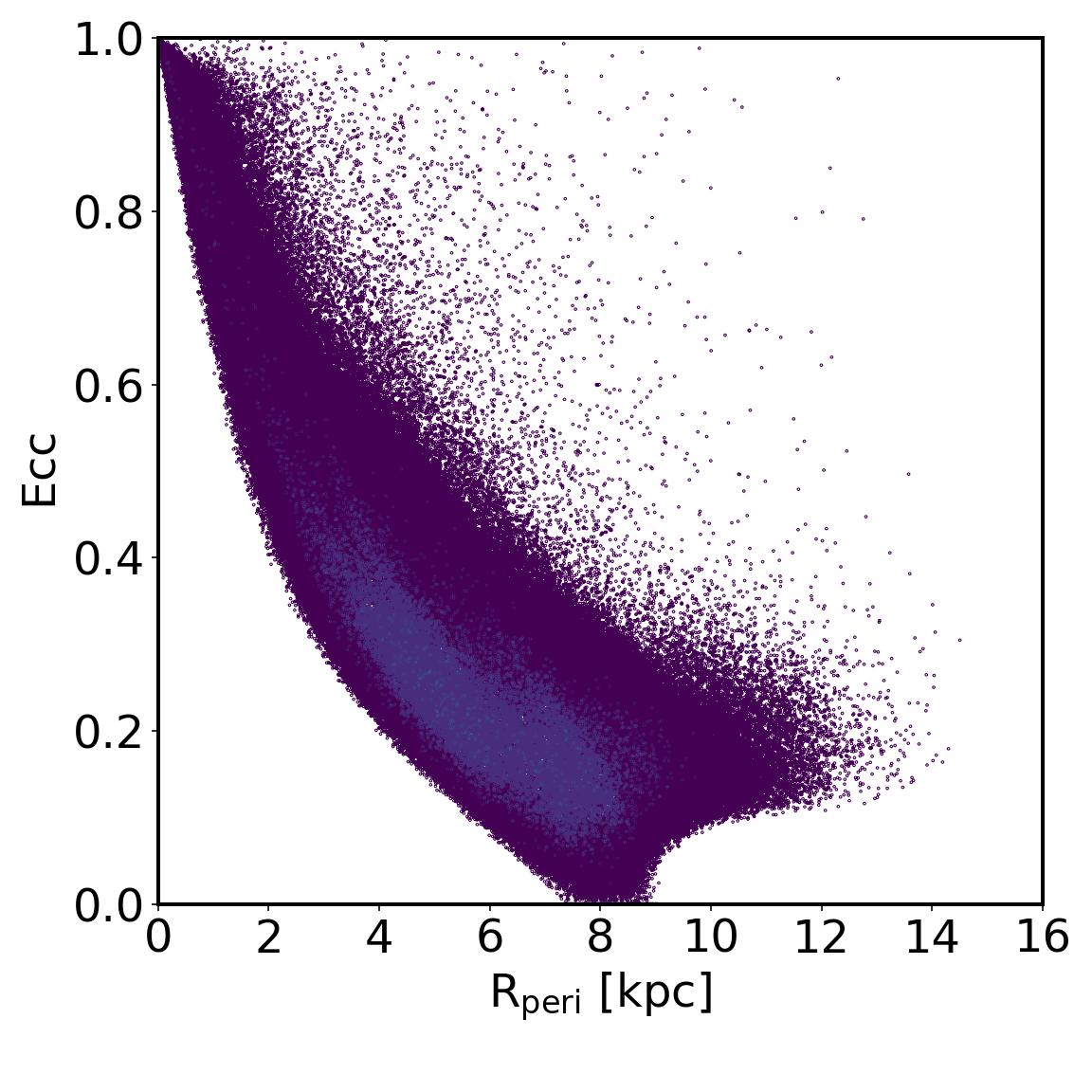}
     \end{subfigure}
     \begin{subfigure}[b]{0.48\textwidth}
         \centering
         \includegraphics[width=\textwidth]{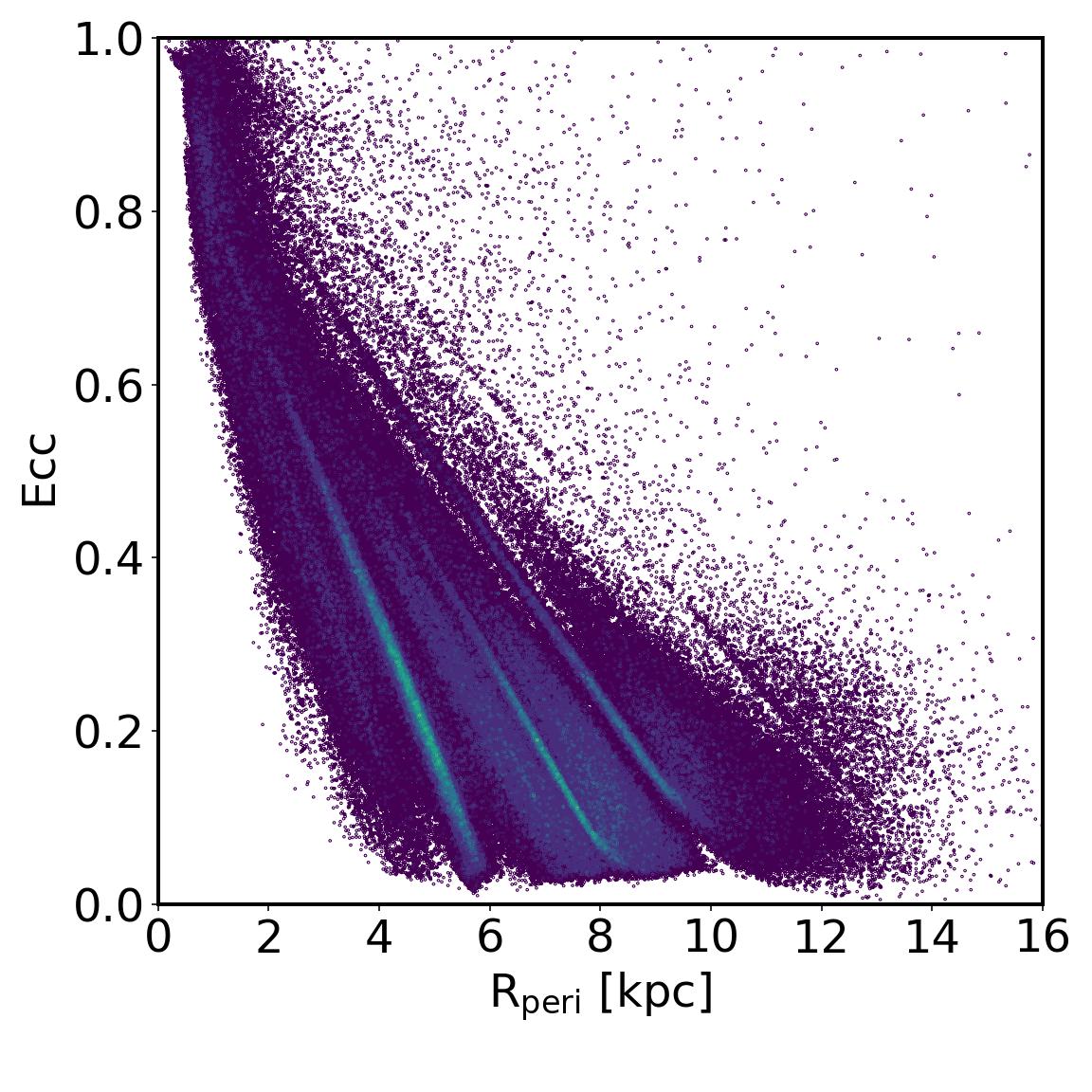}
     \end{subfigure}
        \caption{Density distribution of the full sample in the $R_{peri}-ecc$ plane computed from orbits generated with \textsc{AGAMA}. \textit{Left panel}: the orbital parameters have been recovered using the \textsc{AGAMA} built-in routine \texttt{agama.potential.Rperiapo}. \textit{Right panel}: the orbital parameters have been recovered directly from the orbital history of each tracer.}
        \label{fig:agama}
\end{figure*}

To be sure that our results are model and orbit integrator agnostic, we performed a sanity check running an orbit integration with \textsc{AGAMA} on our sample. To ensure a good enough qualitative comparison we selected for this test the barred MW potential from \citet{2024A&A...692A.216H}, which includes the bar model of \citet{2022MNRAS.514L...1S}, an analytical approximation of the one from \citet{2017MNRAS.465.1621P}. We analysed the results of the \textsc{AGAMA} integration with two different pipelines. First recovering the orbital parameters with the built-in routine \texttt{agama.potential.Rperiapo}\footnote{\href{https://arxiv.org/pdf/1802.08255}{\textsc{AGAMA} reference documentation}} (left panel of Fig.~\ref{fig:agama}) and then by directly deriving them from the recorded orbital history of each tracer (right panel of Fig.~\ref{fig:agama}), as it is done in \textsc{OrbIT}.

The principal and most important difference of the two methods is that the \textsc{AGAMA} built-in function recovers the orbital parameters approximating an axisymmetric potential even when using a barred one, thus erasing from the parameter space the effect of the bar (the overdensities are not thus not present in this case).

The test with the direct orbital parameter recovery (right panel of Fig.~\ref{fig:agama}) confirmed the presence of the overdensities in the orbital parameters phase spaces despite the differences in the underlying potential models assumed and independently from the orbital integration tool used.

\begin{figure}
\centering
\includegraphics[width=\hsize]{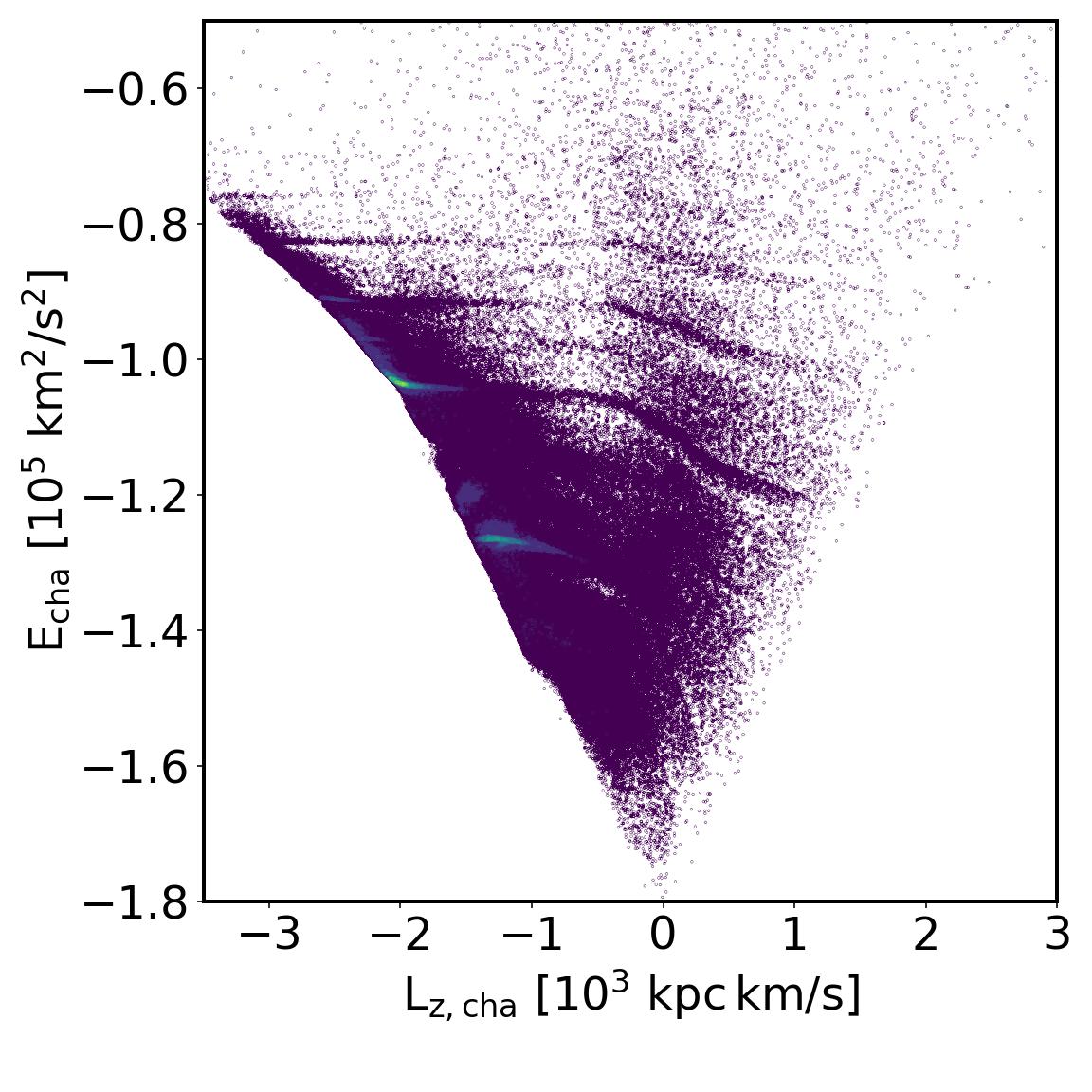}
\caption{Density distribution of the sample stars in $L_{z,cha}-E_{z,cha}$ plane. The diagonal overdensities seen in Fig.~\ref{fig:periecc} appear as almost straight horizontal lines in prograde space ($L_{z,cha} < 0$) and turn at an angle when entering the retrograde region of space ($L_{z,cha} > 0$).}
    \label{fig:lechar}
\end{figure}

\subsection{Resonant trapping of tracers}\label{track}
Having confirmed that the detected overdensities are not caused by systematic or methodological errors and are robust when accounting for uncertainties, we tracked them across the available dynamical parameter spaces to further characterise them. \citet{2015MNRAS.451..705M, 2021MNRAS.506.4687M} had shown how the resonant trapping of orbits operated by the bar would cause clear quasi-horizontal tracks in the $L_{z,cha}-E_{cha}$ plane. This phase space of characteristic angular momentum and energy is analogous of the classical $L_z-E_{tot}$ plane and is used to study quasi-conserved quantities in time-varying potentials (where the IoMs are not conserved). Figure~\ref{fig:lechar} shows the $L_{z,cha}-E_{cha}$ plane for our sample, with the resonant loci tracks as quite clear overdensities.

\begin{figure}
\centering
\includegraphics[width=\hsize]{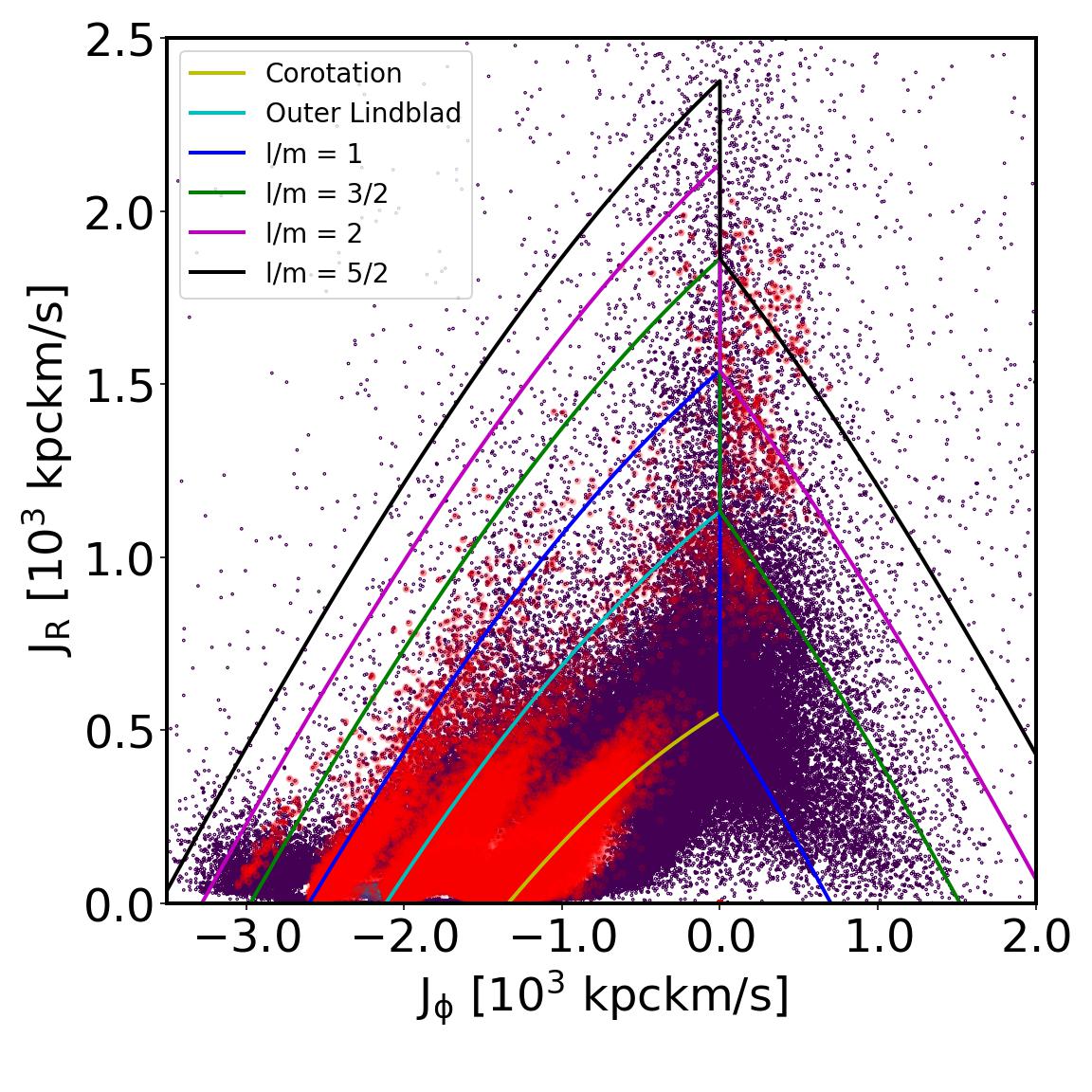}
\caption{The density distribution is made of the stars from our full sample on the $J_{\phi}-J_R$ plane, the red points are the stars on the overdensities in Fig.~\ref{fig:periecc} and Fig.~\ref{fig:lechar}, the coloured solid lines are the theoretical predictions for the loci of the different resonances at specific values of the $l/m$ ratio (see the legend).}
    \label{fig:jphijr}
\end{figure}

Quickly summarising some of the results of the theory of spiral structures and resonances \citep{1953StoAn..17....6L, 1961StoAn..21....8L, 1964ApJ...140..646L, 1966PNAS...55..229L, 1970ApJ...160..113C}, resonant trapping happens when the orbital frequencies of the tracers and the frequency of the perturbation inducing the resonances satisfy a condition of the form:

\begin{equation}
    m(\Omega_{\phi}-\Omega_p)+l \ \Omega_R = 0 \ ,
\end{equation}

where $\Omega_R$ and $\Omega_{\phi}$ are, respectively, the frequencies of radial and azimuthal motion, $\Omega_p$ is the pattern speed of the bar (the perturber) and $l,m$ are integers.

Depending on the ratio $l/m$, several resonant trapping loci can be defined, starting with the most famous corotation (CR), inner and outer Lindblad resonances (I/OLR). To the authors' knowledge, the only potential for which there is a full analytic description of the resonances, down to the prediction of their loci in action space, is the isochrone model \citep{1959AnAp...22..126H, 2008gady.book.....B}. This is because, in this type of potential, the Hamiltonian, action-angle variables, and the associated frequencies can be written as analytic functions of each other \citep{1990MNRAS.244..111E}, thus allowing the direct application of the resonant conditions \citep{1979MNRAS.187..101L, 1996MNRAS.278..395E, 2008gady.book.....B}. Despite the isochrone model being a simplified approximation of the actual potential of the MW \citep[especially in the very central regions, see Fig. 1 from ][]{2024MNRAS.532.4389D}, it remains the only model to produce an analytical prediction for the resonant loci. $J_R$ can then be written as a function of $J_{\phi}$ and the ratio $l/m$ as follows:

\begin{equation}
    \begin{split}
        J_{R,(l,m)} = \bigg[\frac{(GM_{isoc})^2}{\Omega_p}\bigg]^{1/3} \\
        \bigg[\frac{1}{2}\bigg(1+\frac{L}{\sqrt{L^2+4GM_{isoc}b_{isoc}}}\bigg){\rm sgn}(J_{\phi})+\frac{l}{m}\bigg]^{1/3} \\
        -\frac{1}{2}\bigg(L+\sqrt{L^2+4GM_{isoc}b_{isoc}}\bigg)
    \end{split}
\end{equation}
where $L = J_{\theta}+|J_{\phi}|$, $\Omega_p$ is the same used in the orbit integration, the scale length of the isochrone potential is $b_{isoc}=2.95$ kpc and its mass $M_{isoc}=2.23 \times10^{11}M_{\odot}$. While numerical methods could possibly provide more accurate results, the isochrone model provides an exact analytical solution for the resonant loci that is enough for the limited qualitative comparison that we perform in action space and we reserve exploration of other methods for future work.
To compare with the isochrone model, we tracked the stars in the detected overdensities to the $J_{\phi}-J_R$ plane (Fig.~\ref{fig:jphijr}) where we can identify the theoretical loci of resonances induced by the bar rotating with pattern speed $\Omega_p$. From the figure it is evident that, even if the theoretical approximation used is quite distant from the underlying potential of our orbit integration, the overdensities track well the loci of the bar-induced resonances.

\begin{figure}
\centering
\includegraphics[width=\hsize]{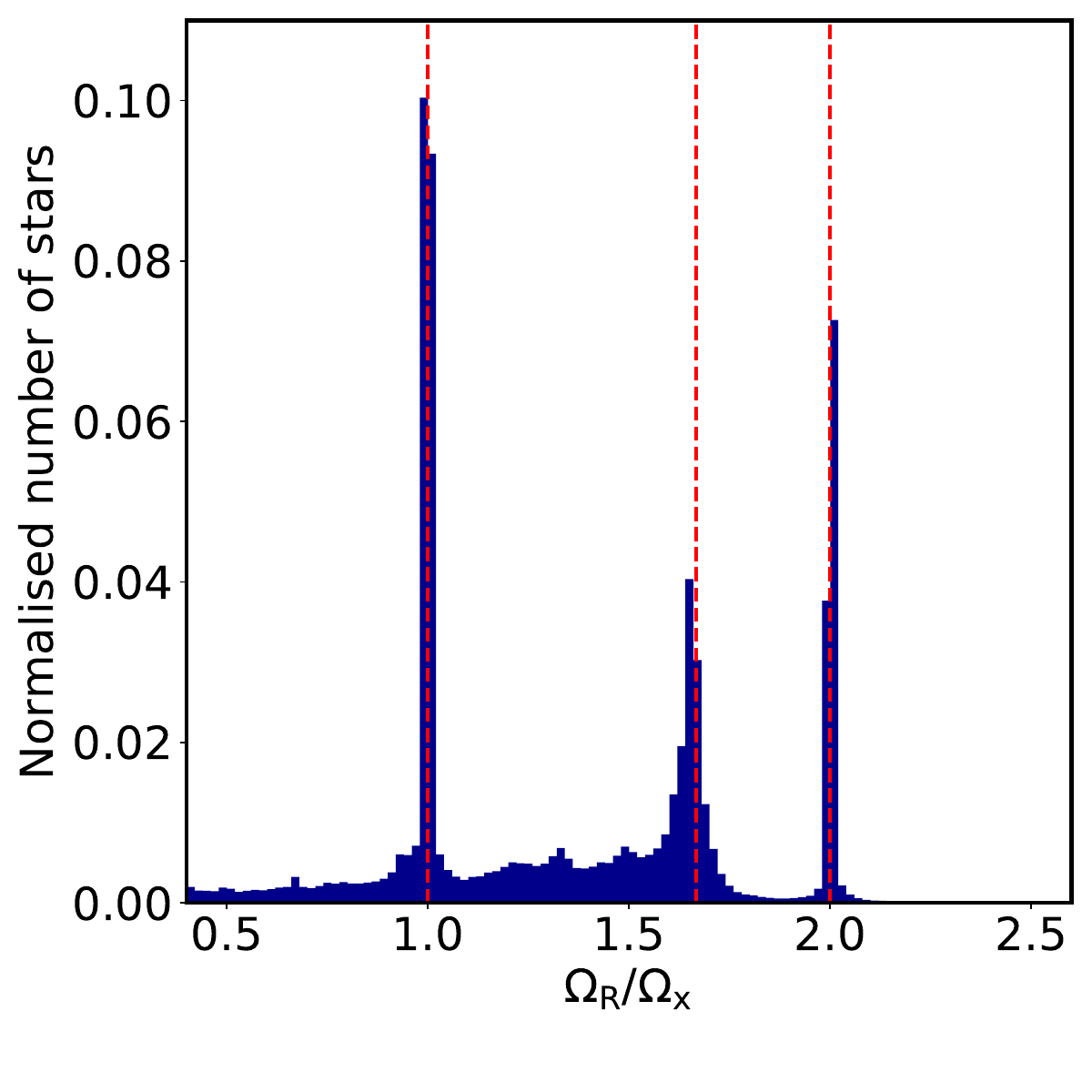}
\caption{Distribution of orbital frequency ratios $\Omega_R/\Omega_x$ for our full sample, the red dashed lines mark the resonant loci at ratios equal to $1, \ 5/3 \ {\rm and} \ 2$.}
    \label{fig:freq}
\end{figure}

As a further and final test, we computed the orbital frequencies of the full sample using \texttt{SuperFreq} \citep{2015ascl.soft11001P} on the 3D position of each tracer at each timestep in the reference frame corotating with the bar. The frequency analysis technique \citep{1990Icar...88..266L, 1993PhRvL..70.2975D, 2008gady.book.....B, 2009MNRAS.394.1605H} uses Fourier transforms of the time series of the coordinates of an orbit to derive its main frequencies and ascertain the degree of regularity or chaoticity. \texttt{SuperFreq} uses a Fast Fourier transform method to derive the fundamental frequencies of an orbit, one for each axis of motion (i.e. $\Omega_x$ for the frequency of oscillations along the $x$ axis). The theoretical expectation from frequency analysis is that bar-supporting orbits would have $\Omega_R/\Omega_x = n$ with $n$ being a specific integer or fraction \citep{2008gady.book.....B, 2015MNRAS.450L..66P, 2021A&A...656A.156Q}. Given the observational errors present in real data, it is customary to accept a $\pm0.1$ tolerance on the frequency ratio \citep{2015MNRAS.450L..66P, 2021A&A...656A.156Q}. For a sample without resonant loci, the distribution of frequency ratios outside the physical size of the bar should be featureless and almost flat, instead we find that the stars on the overdensities in $R_{peri}-ecc$ (i.e. at any radius) inhabit conspicuous peaks in the distribution of $\Omega_R/\Omega_x$ at $n=1,5/3,2$ (Fig.~\ref{fig:freq}).

To summarise, we provide three different criteria, to select tracers under suspicion of being on a resonant orbit. We have shown the resonant loci in $R_{peri}-ecc$ space, and those in $L_{z,cha}-E_{cha}$. If a tracer ends up directly on the loci in both parameter spaces and has a frequency ratio $\Omega_R/\Omega_x$ in one of the peaks, it is considered to be trapped on a resonant orbit. Given the width of the overdensities in Figures ~\ref{fig:periecc} and ~\ref{fig:lechar} and to allow for uncertainties in the estimation of the parameters, we establish strict tolerances on each criterion used. For the four tracks in $R_{peri}-ecc$ space (Fig.~\ref{fig:periecc}), we define tracers as trapped if they have an $ecc$ within $0.03, \ 0.02, \ 0.015 \ {\rm and} \ 0.01$ of each overdensity, going from the leftmost to rightmost one and taking into account their different thickness. For the resonant loci in $L_{z,cha}-E_{cha}$ space (Fig.~\ref{fig:lechar}) a tracer is trapped if it has $E_{cha}$ within $0.01 \ 10^5 \ {\rm km}^2 / {\rm s}^2$ of the tracks. For the frequency ratio it is customary to accept a $\pm0.1$ tolerance \citep{2015MNRAS.450L..66P, 2021A&A...656A.156Q}.
The analytic description of the \textquote{selection boxes} resulting from our criteria (and the loci themselves) are reported (and shown) in Appendix~\ref{re_loc}.

\section{Interpretation of known substructures}\label{subs}
Clustering of tracers in dynamical phase spaces is one of the premiere techniques for discovering and identifying substructures (be they in-situ or accreted merger remnants) among the MW stellar and globular cluster (GC) populations. From the first systematic study by \citet{2019A&A...630L...4M} and many more recent works \citep[i.e.][De Leo et al. submitted]{2019MNRAS.488.1235M, 2019A&A...631L...9K, 2020ApJ...901...48N, 2023MNRAS.520.5671H}, this information has often been complemented with chemical analysis to provide classification for the MW stars and GCs \citep{2020MNRAS.493.3363H, 2022MNRAS.513.4107C, 2023MNRAS.520.5671H, 2024MNRAS.528.3198B, 2024A&A...684A..37C, 2024A&A...687A.201B, 2024A&A...691A.226C}. In the previous sections, we have shown that resonant trapping induced by the rotating bar can produce overdensities of tracers in dynamical parameter spaces and thus can lead to mistaken identification of substructures. Our findings are in good agreement with the recent work of \citet{2025A&A...700A.240W} and \citet{2025MNRAS.542.1331D}, and our analysis highlights an easy and quick method to identify resonant loci directly in the orbital parameter spaces.

We investigated the stellar population of several MW substructures to check if they might be partially or totally explained by bar-induced resonant trapping.

To this end we ran \textsc{OrbIT} on the sample of stars attributed to many substructures identified as overdensities in phase space and tentatively attributed to relics of past merger events. We included in our analysis the selection samples of some of the overdensities identified by \citet{2022A&A...665A..57L, 2022A&A...665A..58R, 2023A&A...670L...2D} and the originally selected samples of candidate members for Nyx \citep{2020NatAs...4.1078N}, LMS-1 \citep{2021ApJ...920...51M}, and Shakti and Shiva \citep[][and private communication from K. Malhan]{2024ApJ...964..104M}. Where only coordinates and/or identifiers are provided, we crossmatched the catalogues with \textit{Gaia} DR3 \citep{2023A&A...674A...1G} and selected the stars with the usual \textit{Gaia} quality cuts \citep{2021A&A...649A...2L}. After correcting for the zeropoint offset \citep{2021A&A...649A...4L}, we obtained the distances for these samples by inversion of the \textit{Gaia} parallaxes.

By checking roughly how many tracers of a given substructure end up on or near resonant loci, we can gauge the degree of contamination of the substructure by resonances. For any given tracer, inhabiting a resonant orbit and belonging to a specific structure are not mutually exclusive conditions. The problem arises when a candidate structure cannot easily be distinguishable from nearby or overlapping (in chemo-dynamical parameter spaces) structures if not for the fact of being an overdensity. In cases like this, it is important to determine how much of a density boost was due to resonant trapping and if the candidate structure would have been detected anyway without that contribution.

\begin{figure}
\centering
\includegraphics[width=\hsize]{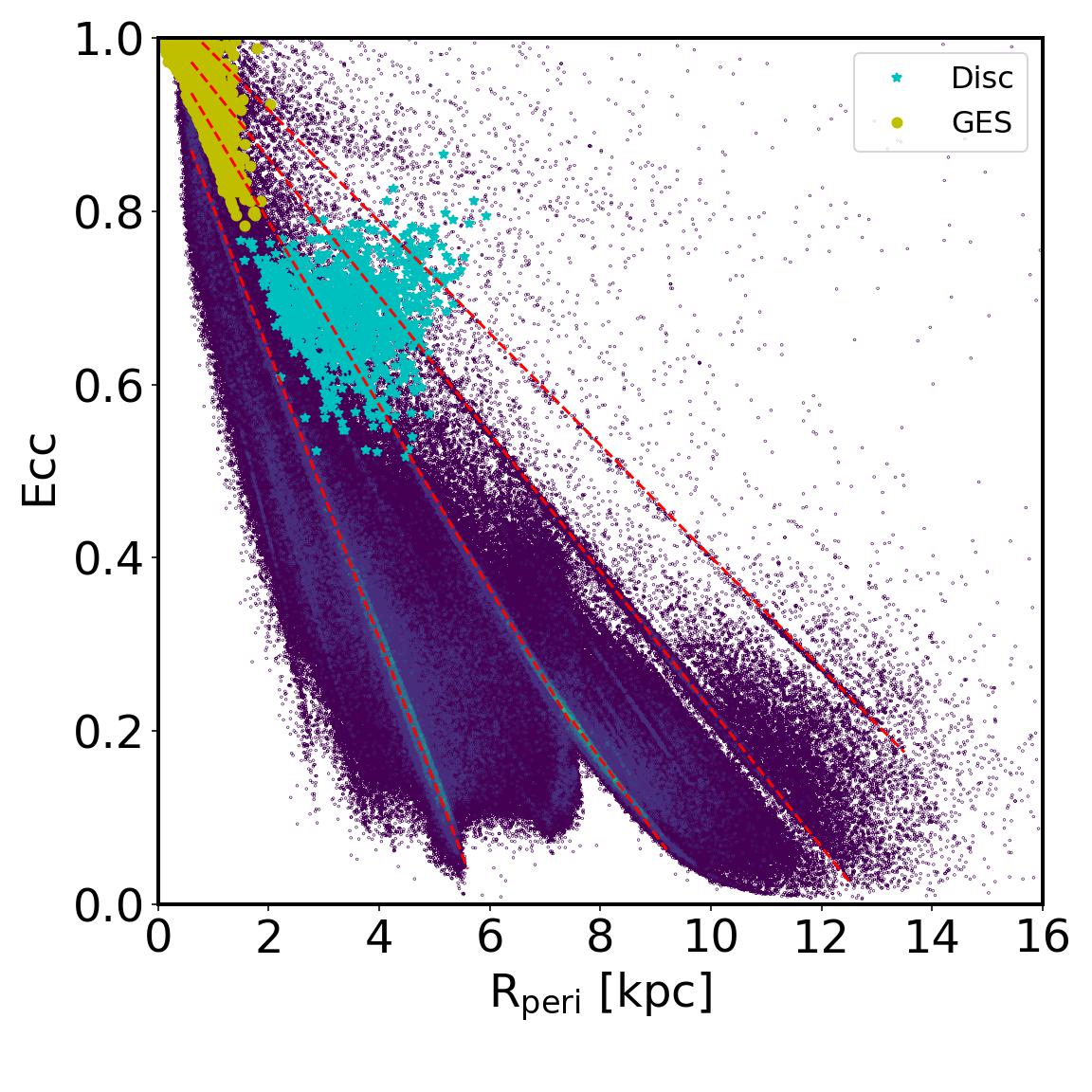}
\caption{The distribution of tracers belonging to the thick disc (cyan stars) and GES (yellow circles), overlaid on our main sample (underlying colour map), in the $R_{peri}-ecc$ plane. The red dashed lines identify the resonant trapping loci.}
    \label{fig:survivors}
\end{figure}

\begin{figure}
\centering
\includegraphics[width=\hsize]{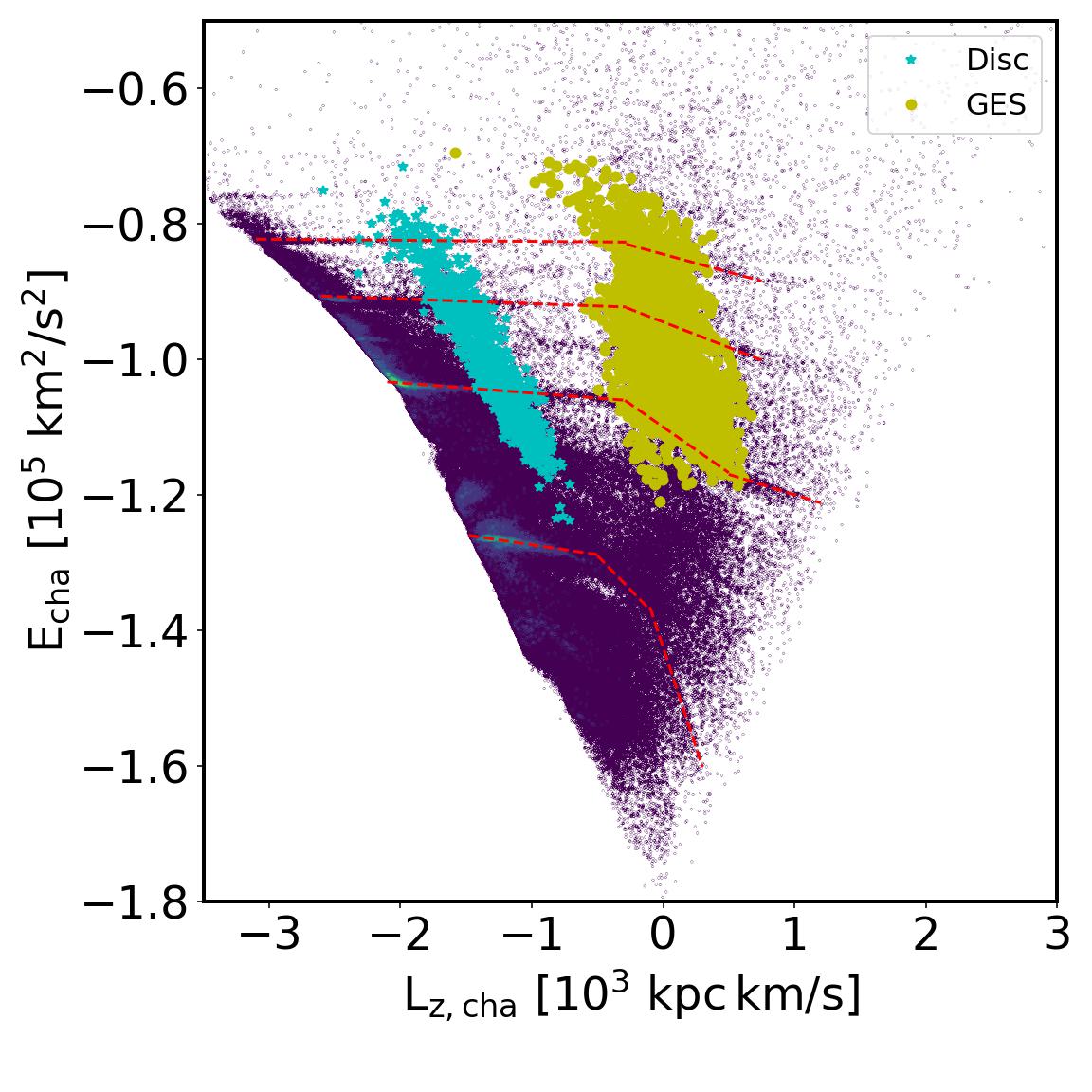}
\caption{The distribution of tracers belonging to the thick disc (cyan stars) and GES (yellow circles), overlaid on our main sample (underlying colour map), in the $L_{z,cha}-E_{cha}$ plane. The red dashed lines identify the resonant trapping loci.}
    \label{fig:survivors_cha}
\end{figure}

Fig.~\ref{fig:survivors} shows the stellar distribution of the members identified in \citet{2023A&A...670L...2D} of the thick disc (in cyan) and Gaia-Enceladus-Sausage \citep[GES, ][in yellow]{2018ApJ...860L..11K, 2018MNRAS.478..611B, 2018ApJ...863..113H, 2018Natur.563...85H}, superimposed on our main sample on the $R_{peri}-ecc$ plane. Given the large number of tracers belonging to the GES (believed to be the most massive accretion event of the MW) and the disc, we are not surprised to find some of them on resonant loci (especially for the disc, as the orbits are prograde and more planar). Nonetheless, both of these structures have tracers spread around large regions of phase space (even more evident in Fig.~\ref{fig:survivors_cha}) and have distinct and well defined chemo-dynamical identities. The main result of applying our analysis to these structures is verifying that it doesn't mistakenly identify all existing substructures as resonant trapping loci.

Moving on to smaller candidate merger events, we study the relation with the resonances, if any, of the stellar populations of Nyx \citep{2020NatAs...4.1078N, 2020ApJ...903...25N, 2022NatAs...6..866N}, Thamnos \citep{2019A&A...631L...9K}, Shakti and Shiva \citep{2024ApJ...964..104M}, LMS-1/Wukong \citep{2020ApJ...901...48N, 2020ApJ...898L..37Y}, Cluster 3 \citep{2022A&A...665A..57L, 2022A&A...665A..58R} and the Helmi Streams \citep{1999Natur.402...53H, 2019A&A...625A...5K}.

\begin{figure}
\centering
\includegraphics[width=\hsize]{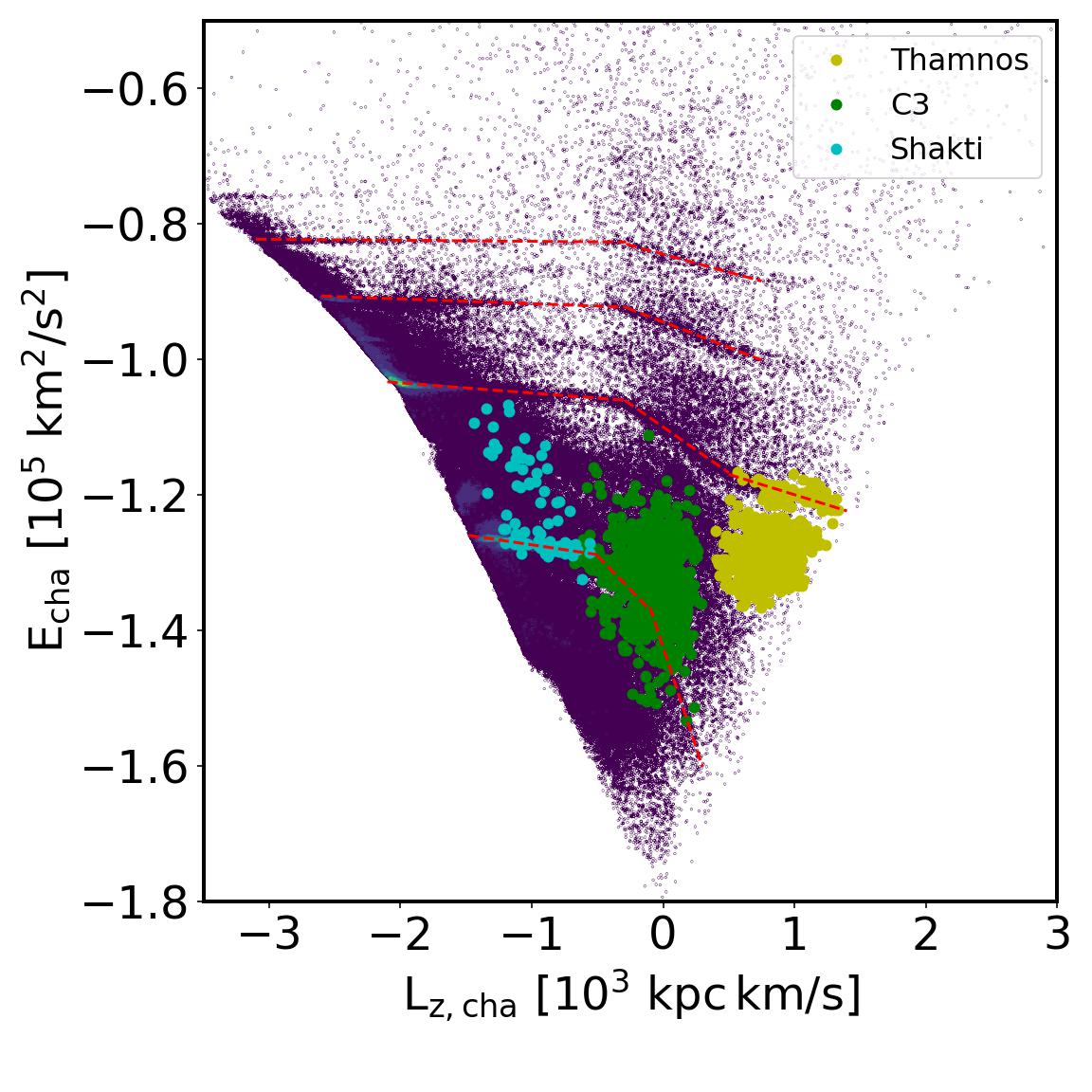}
\caption{The distribution of tracers belonging to Thamnos (yellow points), C3 (green points) and Shakti (cyan points), overlaid on our main sample (underlying colour map), in the $L_{z,cha}-E_{cha}$ plane. The red dashed lines identify the resonant trapping loci.}
    \label{fig:ThamShak_cha}
\end{figure}

\begin{figure}
\centering
\includegraphics[width=\hsize]{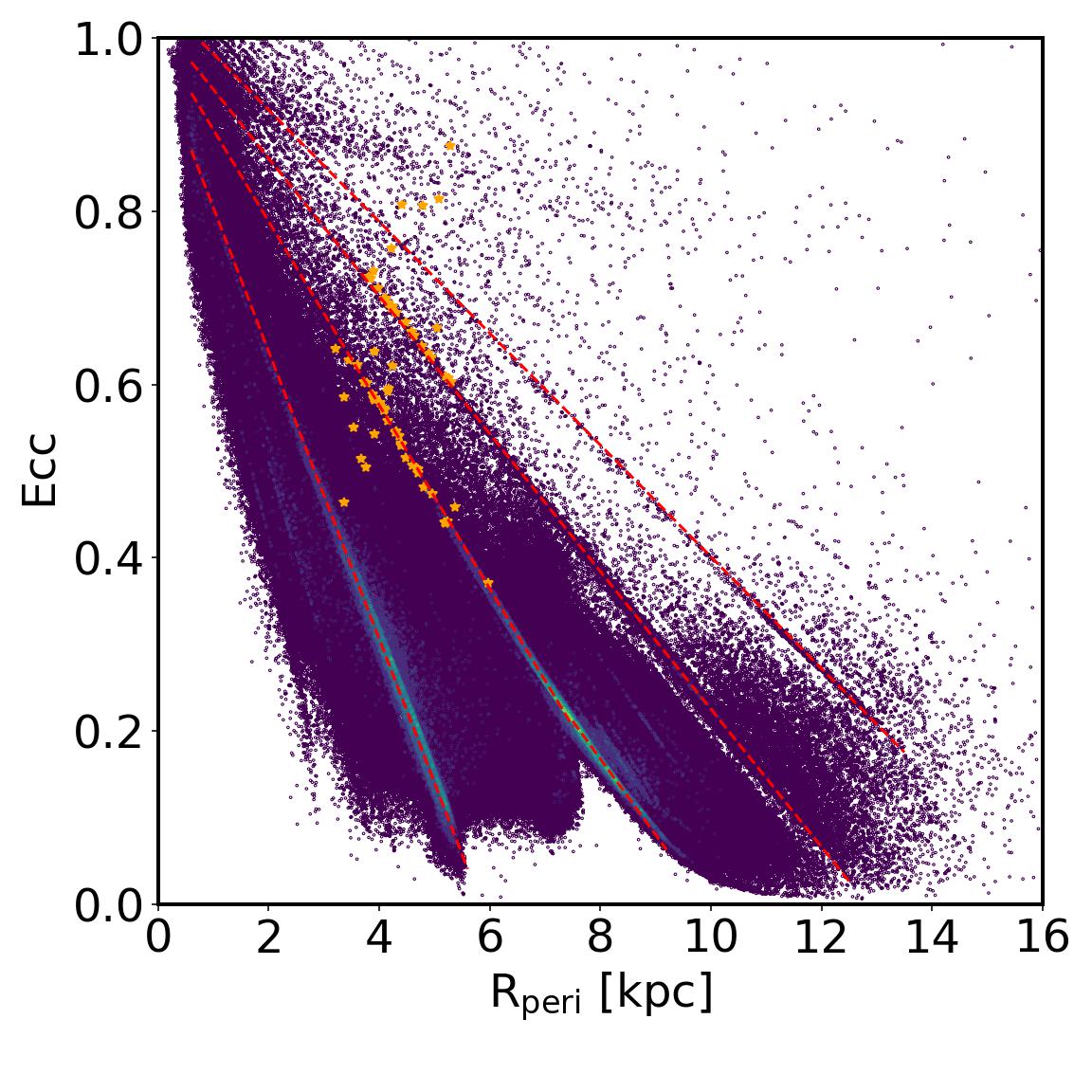}
\caption{The distribution of tracers belonging to Nyx (orange stars), overlaid on our main sample (underlying colour map), in the $R_{peri}-ecc$ plane. The red dashed lines identify the resonant trapping loci.}
    \label{fig:Nyx_periecc}
\end{figure}

\begin{figure}
\centering
\includegraphics[width=\hsize]{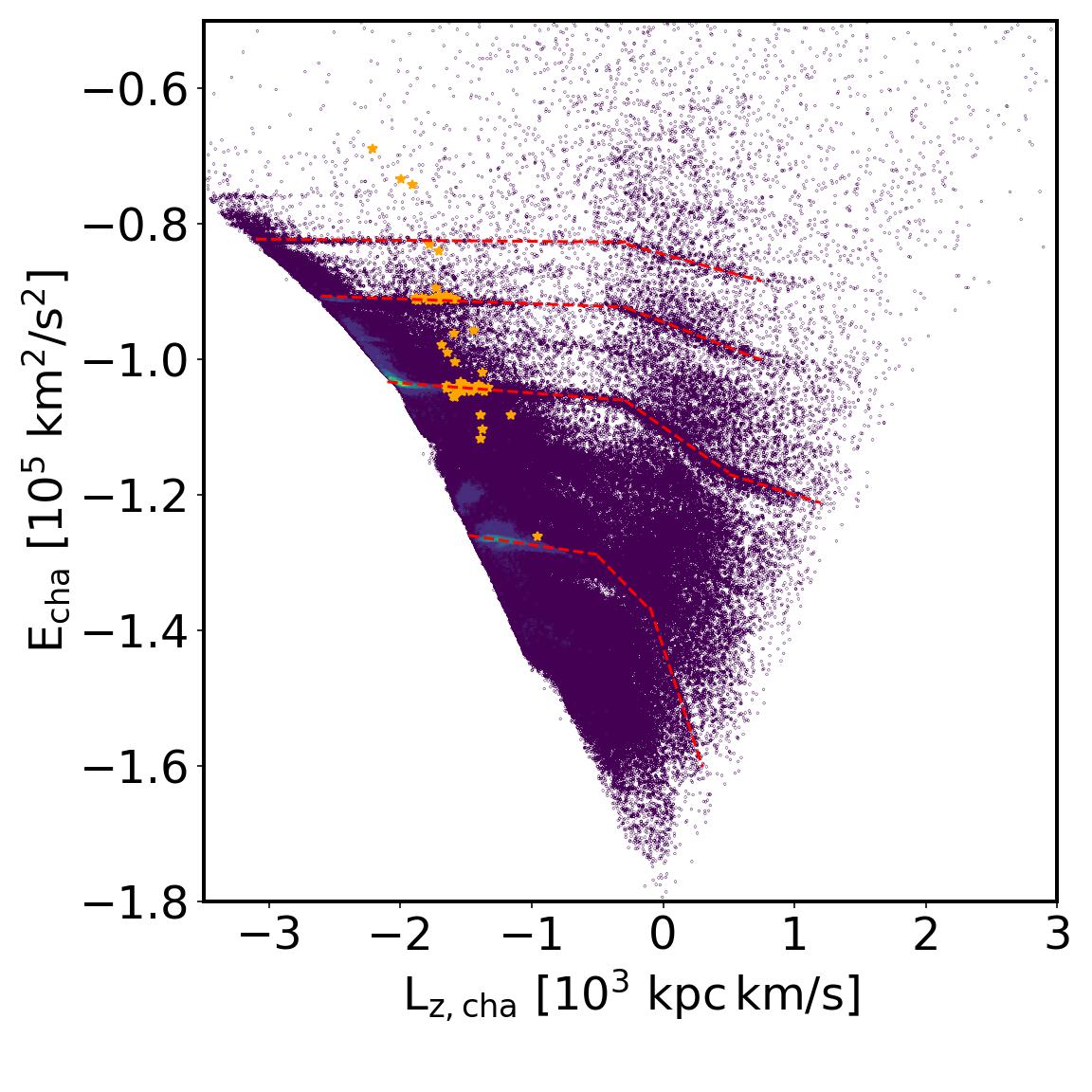}
\caption{The distribution of tracers belonging to Nyx (orange stars), overlaid on our main sample (underlying colour map), in the $L_{z,cha}-E_{cha}$ plane. The red dashed lines identify the resonant trapping loci.}
    \label{fig:Nyx_cha}
\end{figure}

In the cases of Shiva, LMS-1/Wukong and the Helmi Streams, our conservative selection finds a small fraction ($<10\%$) of stars on resonant orbits for each structure. For Cluster 3 (green points in Fig.~\ref{fig:ThamShak_cha}) and Shakti (cyan points in the same figure) we find, respectively, that $16.4\% \ {\rm and} \ 18.5\%$ of their member stars are on resonant orbits. Thamnos (yellow points in Fig.~\ref{fig:ThamShak_cha}), the only purely retrograde structure we analyse, has no stars on resonant orbits according to our strict criteria. Finally, Nyx has the highest number of stars on resonant orbits, with $69.6\%$. As evidenced by the distribution of Nyx stellar population in both $R_{peri}-ecc$ (Fig.~\ref{fig:Nyx_periecc}) and $L_{z,cha}-E_{cha}$ (Fig.~\ref{fig:Nyx_cha}) spaces, most of the stars are on, or very close to, the resonant loci.

\section{Discussion}\label{discu}
We have shown that the bar-induced resonances produce overdensities clearly visible in different parameter spaces and these can be mistaken for genuine substructures of the MW. Among the structures analysed, and particularly for those spanning a wide range of dynamical parameters, it is only natural to find a small fraction of stars within our selection of trapped stars. Even for smaller candidate merger events like Shiva and LMS-1/Wukong, finding a limited contamination by stars trapped on resonances is not particularly concerning, given that their stars have been orbiting inside the MW potential for a very long time. The cases we are most interested in discussing are the structures contaminated to a higher degree, whose discovery could have been mainly driven by the enhanced clustering caused by resonance-trapping.

\subsection{Cluster 3}\label{C3}
This structure \citep{2022A&A...665A..57L, 2022A&A...665A..58R} sits below the GES in the $L_z-E_{tot}$ plane with mixed dynamical signatures and most of its member stars seems to have in-situ chemistry. While in the present work we do not dive into the exact composition, characteristics and ancestry of this group, we have to mention that the region of dynamical space it inhabits has been linked to the MW disc and some other in-situ populations \citep[Aurora and Poor Old Heart][]{2022MNRAS.514..689B, 2022ApJ...938...21M, 2022ApJ...941...45R}. Whatever the nature of the Cluster 3 structure might be, its detection has received a substantial boost from the density enhancement due to the presence of stars trapped on resonances. To be sure of the robustness of our result, we repeat the analysis for the Cluster 3 stars by taking into account the observational errors. For each member star we produce 100 \textquote{clones} with the 6D information extracted from Gaussian distributions having the observed value of each parameter as the mean of said distribution and the observational uncertainties as the standard deviation. Across this expanded sample of Cluster 3 stars, we find $16.8\%$ of them trapped by resonances, confirming our first estimate.

\subsection{Shakti}\label{shakti}
Of the two candidate merger events identified by \citet{2024ApJ...964..104M}, Shakti (the least bound and more energetic one in the original paper) seems to be contaminated by stars trapped near the corotation resonance. As done for the Cluster 3 in Sec.~\ref{C3}, we repeat our analysis by taking into account observational errors and find $17.7\%$ of the population trapped, in line with our first result. These resonant stars populate a very dense clump of Shakti's tracers just below $E_{cha} = -1.2\times10^5 \ {\rm km}^2/{\rm s}^2$ (clearly visible in Fig.~\ref{fig:ThamShak_cha}) and we question if, without this substantial density enhancement, the structure would have been identified. Considering that the chemical properties of Shakti are compatible with the in-situ/disc population and the loci in dynamical space inhabited by its tracers are in almost complete overlap with the disc, we are inclined to agree with the in-situ/disc origin hypothesis proposed in \citet{2024ApJ...964..104M} and supported by the literature \citep{2018ApJ...856L..26M, 2024ApJ...971L...4D}.

\subsection{Nyx}\label{Nyx}
\begin{figure}
\centering
\includegraphics[width=\hsize]{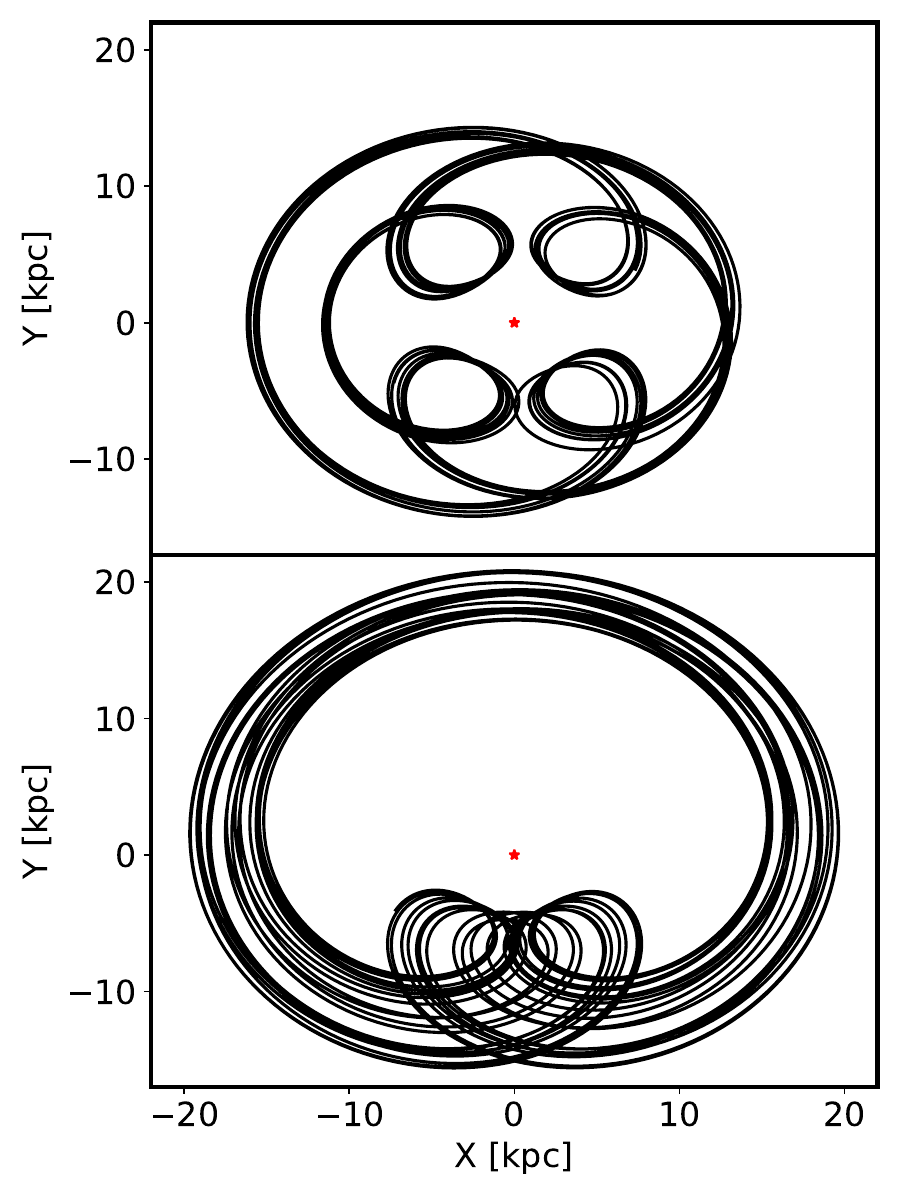}
\caption{In black, two orbits of Nyx stars (representative of the whole sample) on the ${\rm XY}$ plane in the reference frame corotating with the bar, the red star marks the Galactic Centre. It is evident that the stars are resonantly trapped.}
    \label{fig:Nyx_orbs}
\end{figure}

With most of its members stars trapped on the OLR ($l/m = 1/2$) and $l/m = 1$ resonances, it appears clear that Nyx is not a genuine merger event of the MW but rather a structure caused by resonant trapping. To be sure of the robustness of such a strong claim, we repeat the analysis for the Nyx stars by taking into account the observational errors, as done in the previous cases. We find $66.9\%$ of the expanded Nyx sample on trapped orbits, showing that even accounting for the observational uncertainties does not free the Nyx population from the resonances. Visual inspection of the orbits of the stars belonging to Nyx is the final confirmation that they are trapped in resonant loci (Fig.~\ref{fig:Nyx_orbs}). Looking at the chemical evidence available for Nyx stars, recent studies have found disc-like abundance patterns. \citet{2023MNRAS.520.5671H}, examining a sample of APOGEE \citep{2017AJ....154...94M} stars, has found that abundance distributions of the $\alpha$ elements of Nyx members strongly overlaps with the high-$\alpha$ thick disc. Likewise, \citet{2023ApJ...955..129W}, using high resolution spectroscopic data from Keck/HIRES \citep{1994SPIE.2198..362V} and Magellan/MIKE \citep{2003SPIE.4841.1694B}, concluded that the chemical abundances of Nyx members are consistent with the high-$\alpha$ thick disc. These chemical studies put into question Nyx as an independent structure and further strengthen our conclusion that Nyx is an overdensity caused by resonant trapping and not a \textquote{bona fide} merger event.

\subsection{A retrograde outlier: Thamnos}\label{Thamnos}
\begin{figure}
\centering
\includegraphics[width=\hsize]{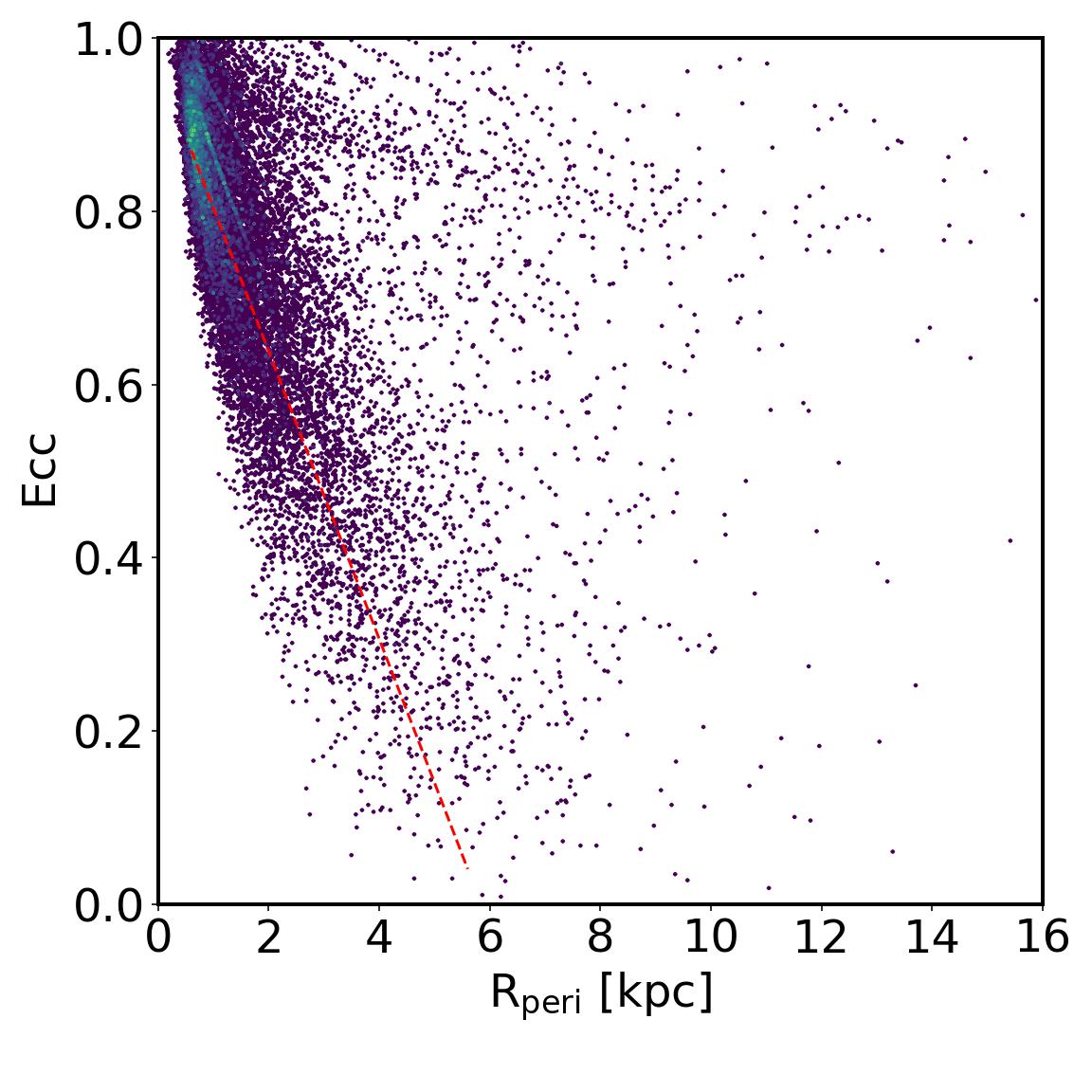}
\caption{Density distribution of the retrograde stars of the sample in $R_{peri}-ecc$ space. The red dashed line sits on top of the first resonant locus identified in Fig.~\ref{fig:periecc} and is clearly different from the quasi-parallel line of the retrograde locus visible on its right.}
    \label{fig:periecc_retro}
\end{figure}

The only purely retrograde structure analysed, Thamnos, appears to be substantially untouched by resonant trapping (with only 1 star passing our threefold criteria out of a sample of 851, or $0.12\%$). This seems to run counter to what is shown in Fig.~\ref{fig:ThamShak_cha} where the highest $E_{cha}$ ridge of Thamnos tracers (the yellow points) appears detached from the rest of the population, sitting on top of the resonance locus around the area of $E_{cha} \simeq -1.2\times10^5 \ {\rm km^2/s^2}$ and $L_{z,cha} \simeq 10^3 \ {\rm kpc \ km/s}$. Given that the resonance loci in the retrograde region ($L_z \equiv J_{\phi} > 0$) seem to always take a different geometry than what is seen in the prograde region (i.e. Figures ~\ref{fig:lechar} and ~\ref{fig:jphijr}), we selected only the retrograde stars in our sample and checked if the resonant loci in $R_{peri}-ecc$ appeared different than what we already found (Fig.~\ref{fig:periecc}). While this retrograde sample is much smaller with $<20000$ stars, it nonetheless allowed us to identify two retrograde resonant loci which are distinct from the ones previously identified in the full sample (completely dominated by the prograde population). Fig.~\ref{fig:periecc_retro} shows the distribution in $R_{peri}-ecc$ space of the retrograde stars in our sample, it is clear that there is no resonant track aligned with the red dashed line (the first resonant locus identified in Fig.~\ref{fig:periecc}) but there is a new one quasi-parallel to the red dashed line on its right.

We parametrise the retrograde resonant loci (see App.~\ref{re_loc}) and repeat our analysis for Thamnos, identifying $8.5\%$ of its members as trapped around the $l/m = 3/2$ resonance. Taking into account the observational errors with the same method used for the other structures gives us $8.2\%$ of the extended sample trapped in resonance, confirming our result. While the portion of Thamnos stars in resonance is quite small, the peculiarity is that it is composed entirely by the stars at the highest energies. Going back to the original identification of this structure, our analysis suggests that \textquote{Thamnos 1} \citep[the most energetic of the two clumps contributing to the structure in][]{2019A&A...631L...9K} might be an overdensity mostly populated by trapped stars. Considering the mounting evidence that Thamnos is heavily contaminated by both in-situ and GES tracers \citep[][]{2022A&A...665A..58R, 2025A&A...698A.277D, 2025arXiv251006332C}, it is important to explore possible alternative explanations for the origin of the stellar population seemingly inhabiting this structure and resonant trapping seems to be able to explain the high energy population.

\section{Conclusions}\label{conclusion}
We have shown that fully accounting for the effect of the bar during the recovery of the orbital parameters allows to identify the loci of bar-induced resonances in different orbital parameter spaces. The tracers trapped on resonances clearly cluster in conspicuous overdensities, easily identifiable, especially in $R_{peri}-ecc$. This unlocks an efficient and expedite way of identifying tracers trapped on resonant orbits already in the space of the orbital parameters ($R_{peri}, \ R_{apo}, \ ecc$) without the need to transition into action space. This transition required several theoretical assumptions and approximations (as regular orbits are extremely rare in real data) that are thus avoided, allowing precise identification of resonances in any potential.

Furthermore, overdensities of tracers generated by the bar-induced resonances and overdensities due to past merger are indistinguishable for clustering algorithms, with the chance of mistaken identification of candidate merger events. We have shown how structures previously identified as mergers of the MW, can be constituted partially or totally by stars tracers on resonant orbits. In particular, cluster 3 and Shakti seem to have received a substantial density enhancement from resonant trapping, while Nyx proves to be an exemplary case of an overdensity which is not a real merger event but instead is caused by stars trapped on resonances. The only purely retrograde structure analysed here, Thamnos, also shows signs of contamination from resonantly trapped stars, which affects is highest energy members (\textquote{Thamnos 1}).

As discussed in Sec.~\ref{track}, the available theoretical framework for the calculation of resonant loci is developed in action space and only offers an approximation of the real, complex situation of the MW. While we have shown the robustness of our empirical method for the identification of resonant structures, we also need to point out the lack of precise theoretical (be they analytical or numerical) predictions of the resonant loci in the dynamical spaces of orbital parameters.

\begin{acknowledgements}

The authors would like to thank the anonymous referee for insightful comments that helped improve and clarify the manuscript. The authors would also like to thank to E. Ceccarelli and E. Dodd for valuable discussions. MDL and AM acknowledge financial support from the project \textquote{LEGO – Reconstructing the building blocks of the Galaxy by chemical tagging} (PI: Mucciarelli) granted by the Italian MUR through contract PRIN2022LLP8TK\_001. BA-T acknowledges support from the ANID Doctoral Fellowship through grant number 21231305.

\end{acknowledgements}

\textit{Software:} this research made use of the Astropy \citep{2013A&A...558A..33A, 2018AJ....156..123A}, Matplotlib \citep{2007CSE.....9...90H} and Numpy \citep{harris2020array} packages.

\section*{Data Availability Statement}

The data underlying this article come from public sources listed in Section ~\ref{data}. The results derived from \textsc{OrbIT} will be made available upon reasonable request.

\bibliographystyle{aa}
\bibliography{orbit}

\begin{appendix}

\section{Comparison of radial action recovery}\label{Jr_comp}
\begin{figure}
\centering
\includegraphics[width=\hsize]{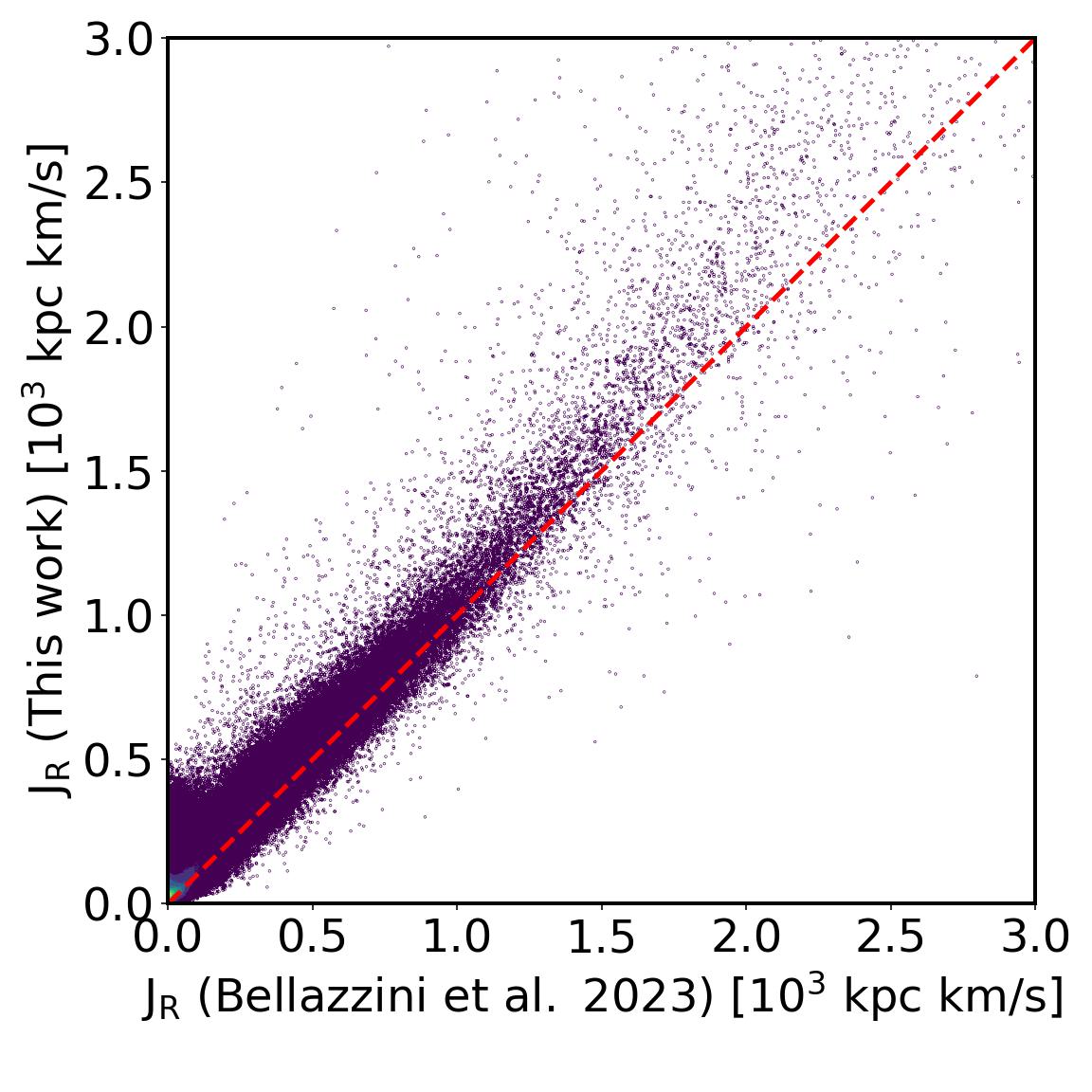}
\caption{Comparison of the $J_R$ recovered in \citet{2023A&A...674A.194B} and in this work, the dashed red line is the equivalence line.}
    \label{fig:Jr_Jr}
\end{figure}

As a sanity check for our numerical method of recovery of the radial action $J_R$, we compare our results with those derived for the same sample in \citet{2023A&A...674A.194B} using a \citet{2017MNRAS.465...76M} potential and the AGAMA action finder built-in routines using the St\"{a}ckel approximation (see the \href{https://arxiv.org/pdf/1802.08255}{\textsc{AGAMA} reference documentation}). Fig.~\ref{fig:Jr_Jr} shows the reasonable agreement of the two different estimations of $J_R$ despite the different methods and underlying potentials used.

\section{Effect of changes in the pattern speed}\label{ps35}
\begin{figure}
\centering
\includegraphics[width=\hsize]{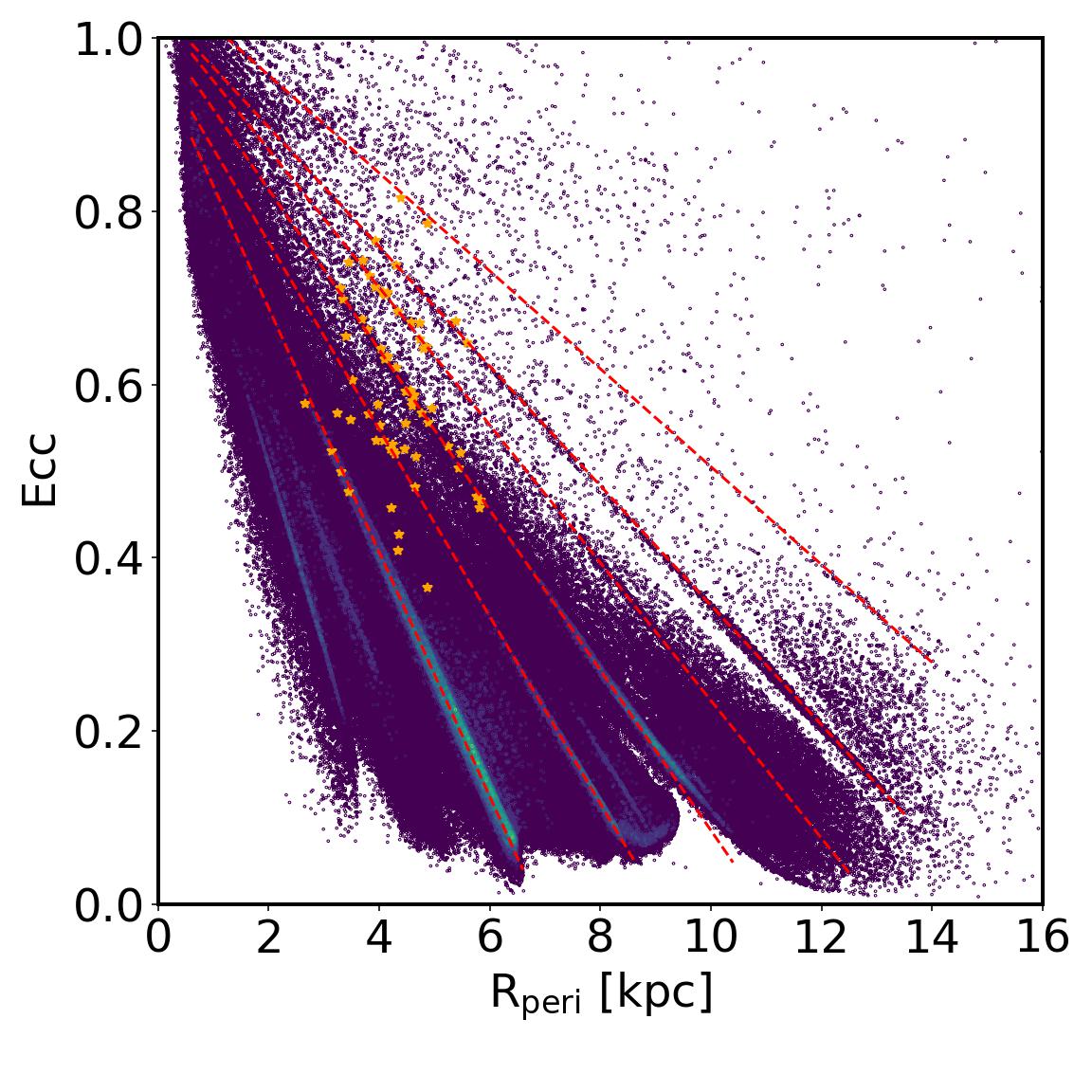}
\caption{Same as Fig.~\ref{fig:Nyx_periecc} but for $\Omega_p = 35 \ {\rm km \ s^{-1} \ kpc^{-1}}$.}
    \label{fig:Nyx35_periecc}
\end{figure}

\begin{figure}
\centering
\includegraphics[width=\hsize]{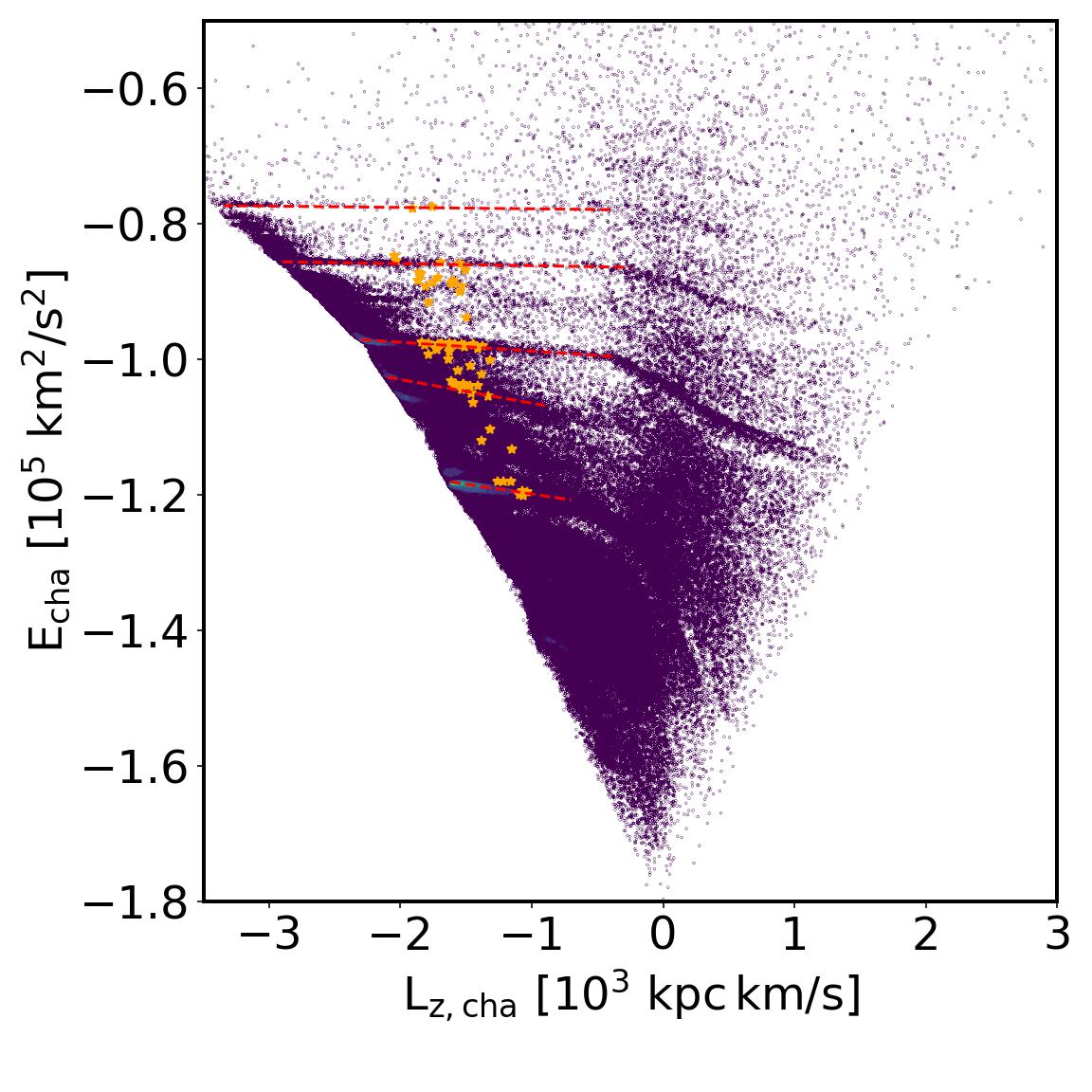}
\caption{Same as Fig.~\ref{fig:Nyx_cha} but for $\Omega_p = 35 \ {\rm km \ s^{-1} \ kpc^{-1}}$.}
    \label{fig:Nyx35_lecha}
\end{figure}

\begin{figure}
\centering
\includegraphics[width=\hsize]{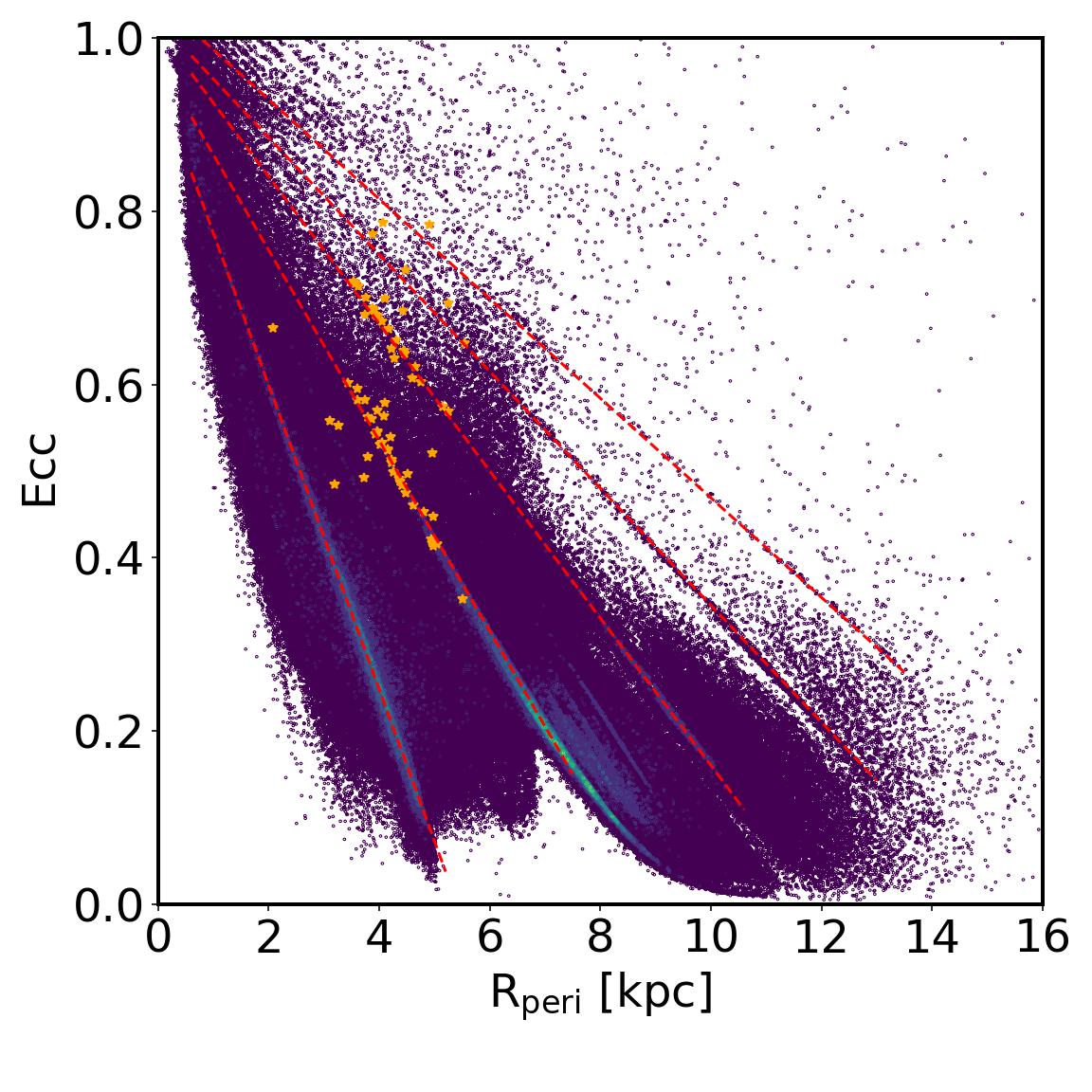}
\caption{Same as Fig.~\ref{fig:Nyx_periecc} but for $\Omega_p = 45 \ {\rm km \ s^{-1} \ kpc^{-1}}$.}
    \label{fig:Nyx45_periecc}
\end{figure}

\begin{figure}
\centering
\includegraphics[width=\hsize]{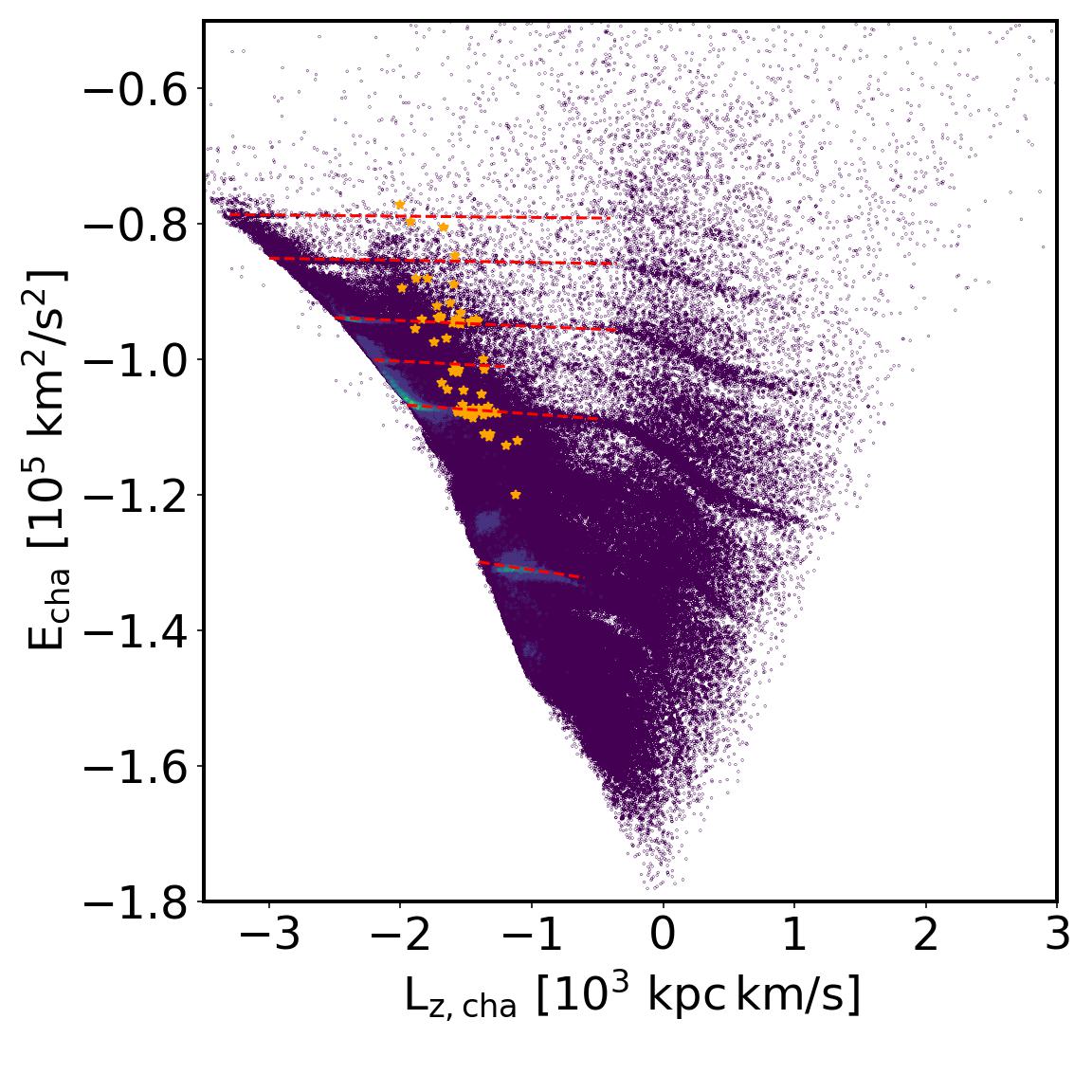}
\caption{Same as Fig.~\ref{fig:Nyx_cha} but for $\Omega_p = 45 \ {\rm km \ s^{-1} \ kpc^{-1}}$.}
    \label{fig:Nyx45_lecha}
\end{figure}

To test the robustness of our method against changes in the pattern speed of the bar, the parameter affecting the resonances the most, we repeated the entire analysis setting $\Omega_p = 35 \ {\rm km \ s^{-1} \ kpc^{-1}}$. We reran \textsc{OrbIT} on the entire sample, and were able to identify the resonance loci in $R_{peri}-ecc \ {\rm and} \ L_{z,cha}-E_{cha}$. As mentioned in Sec~\ref{results}, the loci are slightly moved with respect to the integration at $\Omega_p = 41.3 \ {\rm km \ s^{-1} \ kpc^{-1}}$, used throughout the main text. In order to fully test the robustness of our results, we repeated the analysis of the Nyx sample, the most important finding of this work. Figures ~\ref{fig:Nyx35_periecc} and ~\ref{fig:Nyx35_lecha} show the outcome of this test, with $62.3\%$ of the Nyx member stars ending on the resonant loci. To further prove the robustness of our method and results, we repeated our analysis for the main sample and Nyx members with $\Omega_p = 45 \ {\rm km \ s^{-1} \ kpc^{-1}}$. Again, we were able to identify the resonant loci and show that the majority of Nyx tracers (43 out of 69) are trapped on them (Figures ~\ref{fig:Nyx45_periecc} and ~\ref{fig:Nyx45_lecha}).

\section{Selection of resonant loci}\label{re_loc}
\begin{figure}
\centering
\includegraphics[width=\hsize]{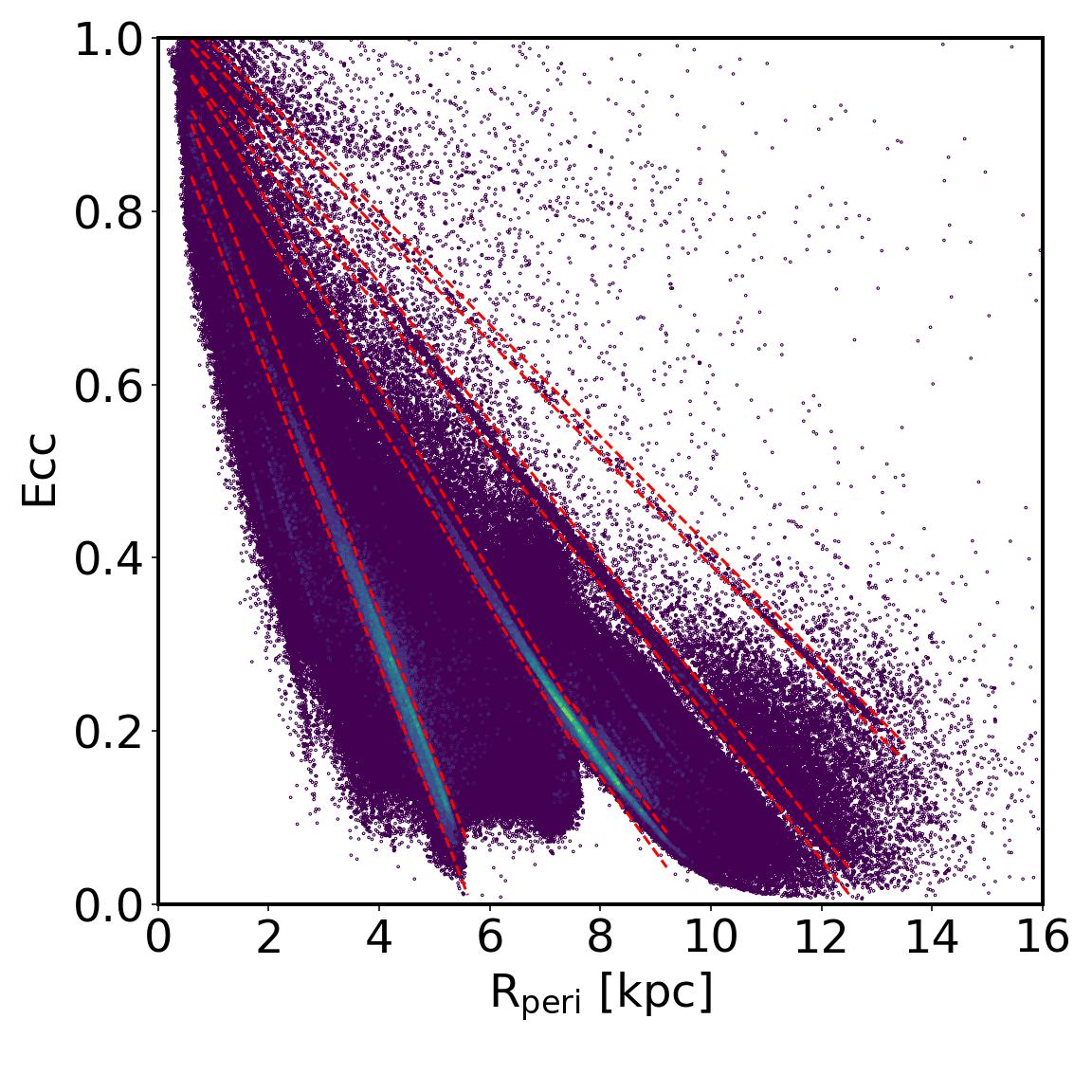}
\caption{Density distribution of the full sample in $R_{peri}-ecc$, the red dahsed lines are the confines of the selections boxes for the resonant loci.}
    \label{fig:re_loc_periecc}
\end{figure}

The resonant loci in $R_{peri}-ecc$ and $L_{z,cha}-E_{cha}$ are identified by the equations reported below and are shown in the Fig.~\ref{fig:re_loc_periecc} and ~\ref{fig:re_loc_lecha}. In $R_{peri}-ecc$ space the resonant loci are parametrised as follows (with the $R_{peri}$ in kpc): \\
- $ecc = 0.97-0.166\cdot R_{peri}$ for $0.6 \leq R_{peri} \leq 5.6$ \\
- $ecc = 1.00-0.106\cdot R_{peri}$ for $0.6 \leq R_{peri} \leq 7.5$ \\
- $ecc = 0.8675-0.0875\cdot R_{peri}$ for $7.5 \leq R_{peri} \leq 9.2$ \\
- $ecc = 1.020-0.0795\cdot R_{peri}$ for $0.6 \leq R_{peri} \leq 12.5$ \\
- $ecc = 1.046-0.0645\cdot R_{peri}$ for $0.6 \leq R_{peri} \leq 13.5$ \\
The intervals marked by the red dashed lines in Fig.~\ref{fig:re_loc_periecc} are obtained by adding to the equation of each locus reported above an offset equal to the tolerances specified in Sec.~\ref{track}: $\pm 0.03, \ \pm 0.02, \ \pm 0.02, \ \pm0.015 \ {\rm and} \ \pm 0.01$.

\begin{figure}
\centering
\includegraphics[width=\hsize]{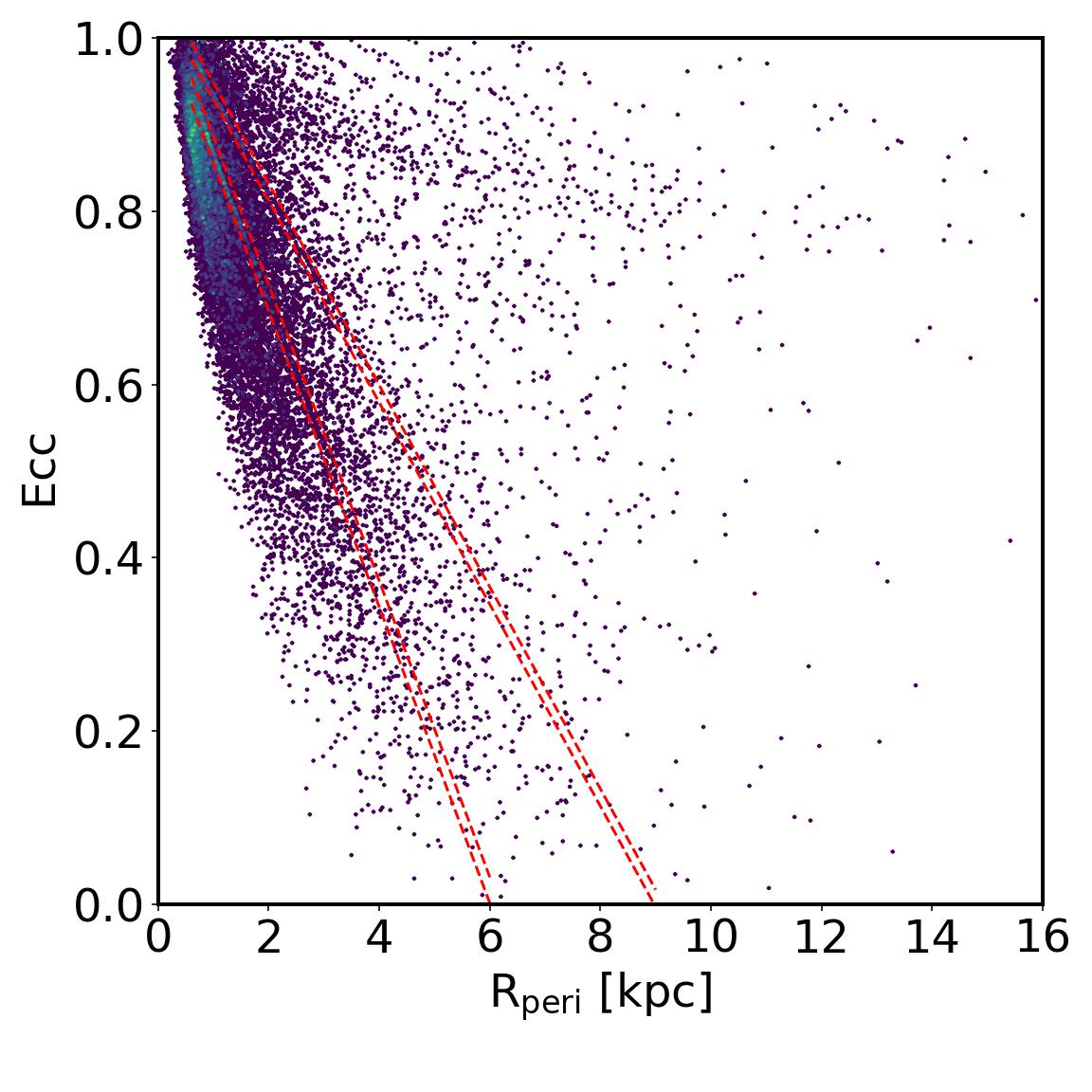}
\caption{As Fig. ~\ref{fig:re_loc_periecc} but for the retrograde stars in the sample.}
    \label{fig:re_loc_periecc_retro}
\end{figure}

\begin{figure}
\centering
\includegraphics[width=\hsize]{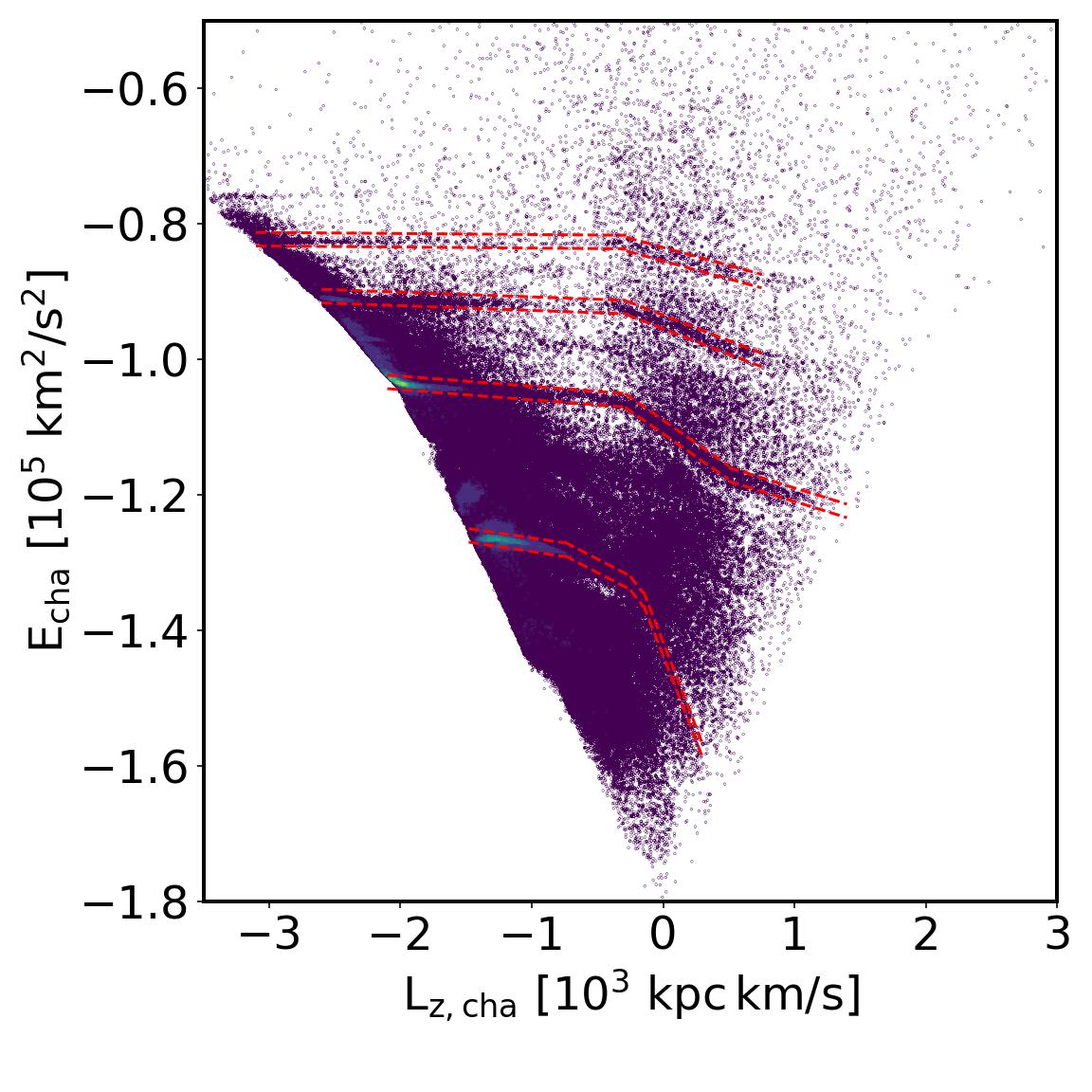}
\caption{As Fig. ~\ref{fig:re_loc_periecc} but in $L_{z,cha}-E_{cha}$ space.}
    \label{fig:re_loc_lecha}
\end{figure}

As explained in Sec.~\ref{Thamnos}, the resonant loci in the retrograde region are different and are parametrised as follows: \\
- $ecc = 1.042-0.171\cdot R_{peri}$ for $0.6 \leq R_{peri} \leq 6.0$ \\
- $ecc = 1.055-0.1165\cdot R_{peri}$ for $0.6 \leq R_{peri} \leq 9.0$ \\
The intervals in Fig.~\ref{fig:re_loc_periecc_retro} are obtained by adding offsets of, respectively, $\pm 0.015, \ {\rm and} \ \pm 0.01$.

In $L_{z,cha}-E_{cha}$ space the resonant loci are parametrised as follows (with $L_{z,cha}$ in $10^3 \ {\rm kpc \ km/s}$ and $E_{cha}$ in $10^5 \ {\rm km^2/s^2}$): \\
- $E_{cha} = -1.303-0.029\cdot L_{z,cha}$ for $-1.48 \leq L_{z,cha} \leq -0.75$ \\
- $E_{cha} = -1.355-0.1\cdot L_{z,cha}$ for $-0.75 \leq L_{z,cha} \leq -0.25$ \\
- $E_{cha} = -1.393-0.25\cdot L_{z,cha}$ for $-0.25 \leq L_{z,cha} \leq -0.15$ \\
- $E_{cha} = -1.4275-0.50\cdot L_{z,cha}$ for $-0.25 \leq L_{z,cha} \leq 0.3$ \\
- $E_{cha} = -1.065-0.015\cdot L_{z,cha}$ for $-2.1 \leq L_{z,cha} \leq 0.29$ \\
- $E_{cha} = -1.10-0.135\cdot L_{z,cha}$ for $-0.29 \leq L_{z,cha} \leq 0.5$ \\
- $E_{cha} = -1.14-0.06\cdot L_{z,cha}$ for $0.5 \leq L_{z,cha} \leq 1.4$ \\
- $E_{cha} = -0.925-0.007\cdot L_{z,cha}$ for $-2.6 \leq L_{z,cha} \leq -0.28$ \\
- $E_{cha} = -0.945-0.075\cdot L_{z,cha}$ for $-0.28 \leq L_{z,cha} \leq 0.75$ \\
- $E_{cha} = -0.8275-0.0015\cdot L_{z,cha}$ for $-3.1 \leq L_{z,cha} \leq -0.28$ \\
- $E_{cha} = -0.845-0.053\cdot L_{z,cha}$ for $-0.28 \leq L_{z,cha} \leq 0.75$ \\
The offset needed to reproduce the intervals in Fig.~\ref{fig:re_loc_lecha} is $\pm0.01 \ 10^5 \ {\rm km^2/s^2}$.

\end{appendix}
\end{document}